\documentclass[10pt,twocolumn,letterpaper]{article}

\usepackage[pagenumbers]{cvpr} 
\usepackage{graphicx}
\usepackage{amsmath}
\usepackage{amssymb}
\usepackage{booktabs}
\usepackage{multirow}
\usepackage{times}
\usepackage{lipsum}

\usepackage{bm,cuted}
\usepackage[ruled,vlined,linesnumbered]{algorithm2e}
\usepackage{graphicx}
\usepackage{tabularx} 
\usepackage[safe]{tipa}
\usepackage{multirow}
\usepackage{xcolor}
\usepackage{dsfont}
\usepackage{upgreek}
\usepackage{url}

\usepackage{pifont}
\makeatletter
\newcommand\HUGE{\@setfontsize\Huge{20}{30}}
\makeatother

\def\eqref#1{equation~\ref{#1}}

\def\1{\bm{1}}

\DeclareMathAlphabet{\mathsfit}{\encodingdefault}{\sfdefault}{m}{sl}
\SetMathAlphabet{\mathsfit}{bold}{\encodingdefault}{\sfdefault}{bx}{n}

\usepackage[accsupp]{axessibility}

\usepackage[pagebackref,breaklinks,colorlinks,bookmarks=false]{hyperref}

\hypersetup{pdfstartview={FitH}}

\usepackage[capitalize]{cleveref}
\crefname{section}{Sec.}{Secs.}
\Crefname{section}{Section}{Sections}
\Crefname{table}{Table}{Tables}
\crefname{table}{Tab.}{Tabs.}

\def\cvprPaperID{5880}
\def\confName{CVPR}
\def\confYear{2023}

\begin{document}

\title{Music-Driven Group Choreography}
\author{Nhat Le$^{1}$, Thang Pham$^{1}$, Tuong Do$^{1}$, Erman Tjiputra$^{1}$, Quang D. Tran$^{1}$, Anh Nguyen$^{2}$\\
{\small $^{1}$AIOZ, Singapore}
{\small $^{2}$University of Liverpool, UK}}

\twocolumn[{%
\renewcommand\twocolumn[1][]{#1}%
\maketitle
\begin{center}
  \centering
  \captionsetup{type=figure}
  \Large
\resizebox{\linewidth}{!}{
\setlength{\tabcolsep}{2pt}
\begin{tabular}{ccccc}
\shortstack{\includegraphics[width=0.33\linewidth]{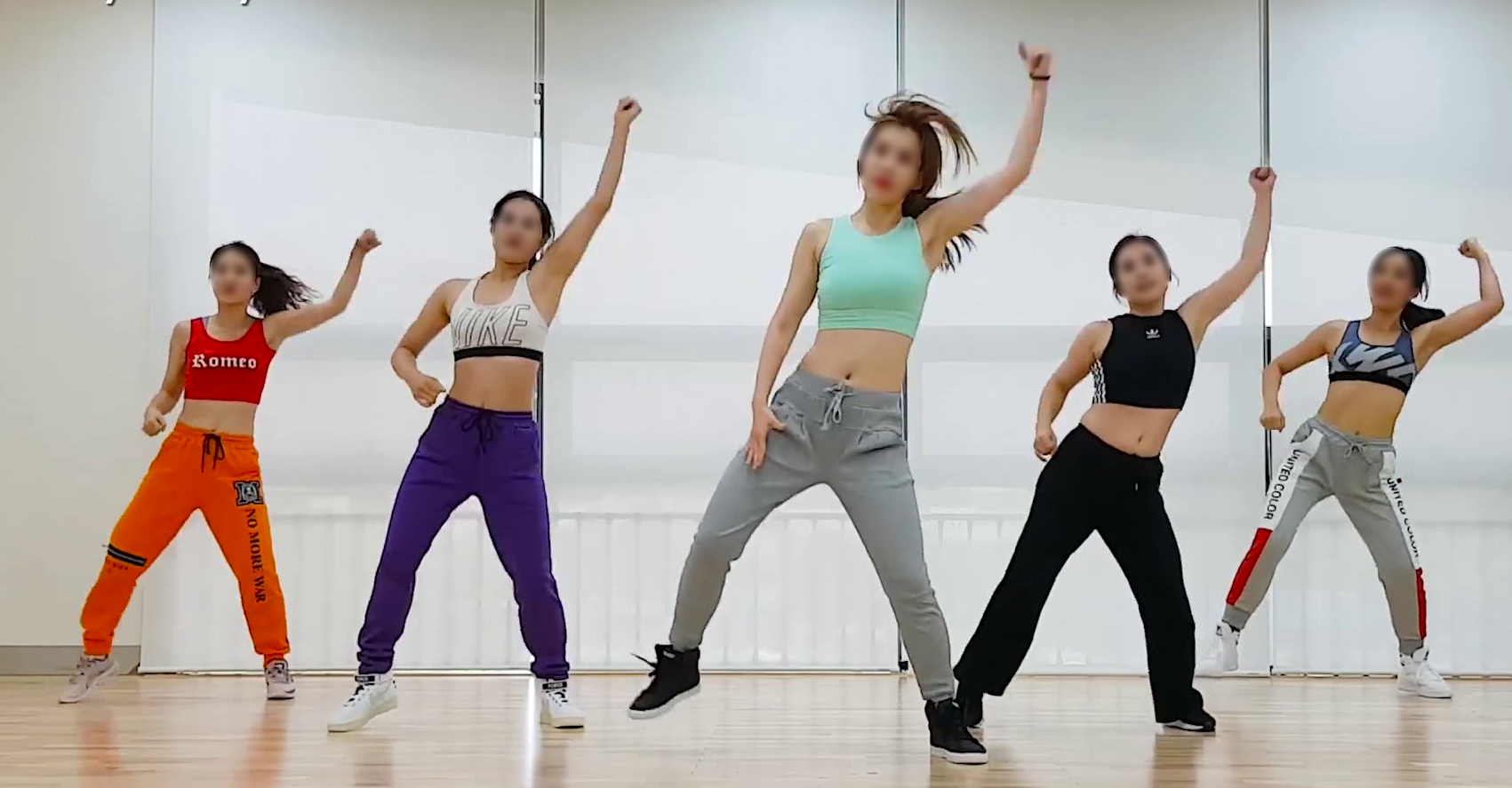}}&
\shortstack{\includegraphics[width=0.33\linewidth]{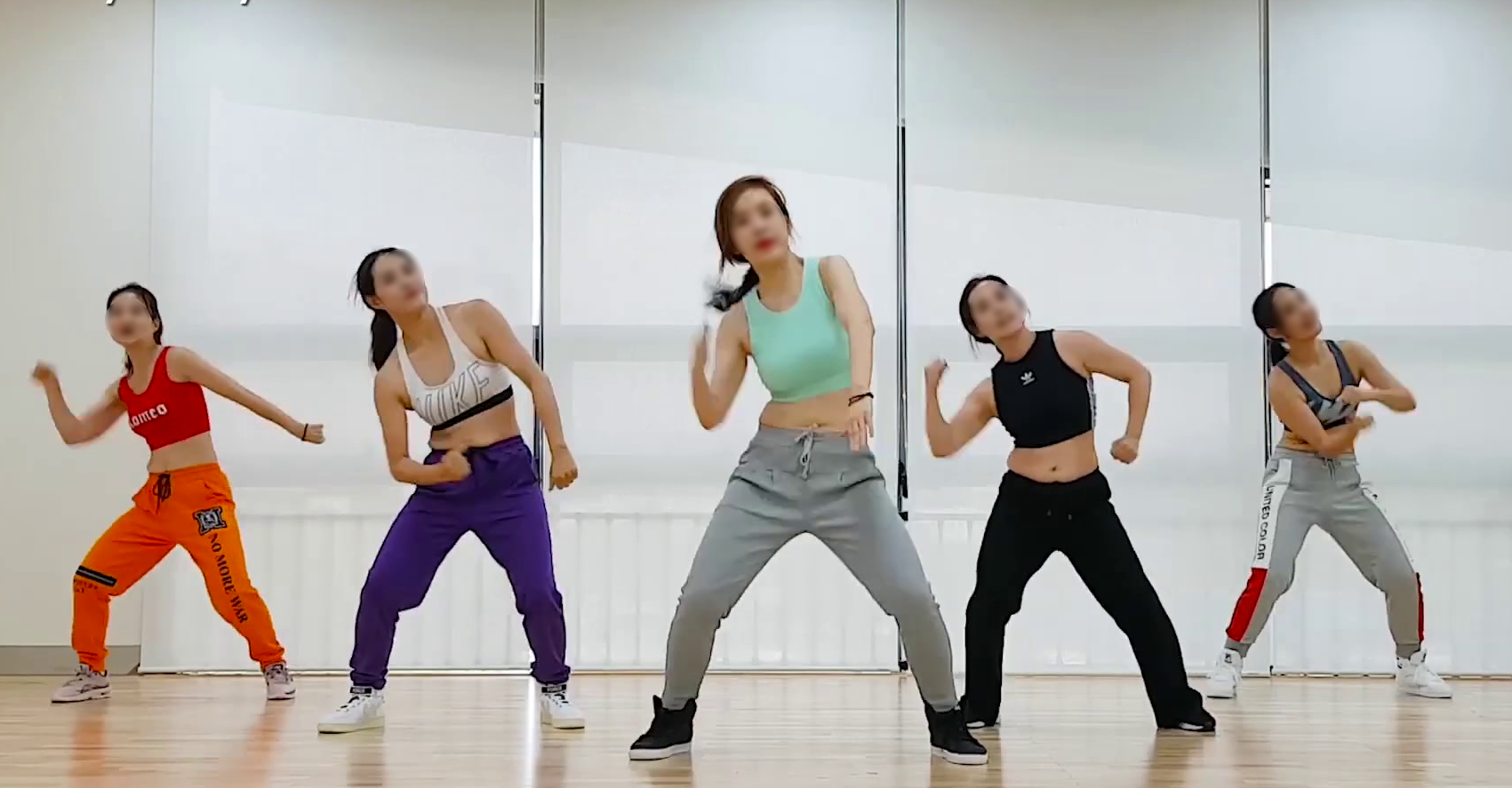}}&
\shortstack{\includegraphics[width=0.33\linewidth]{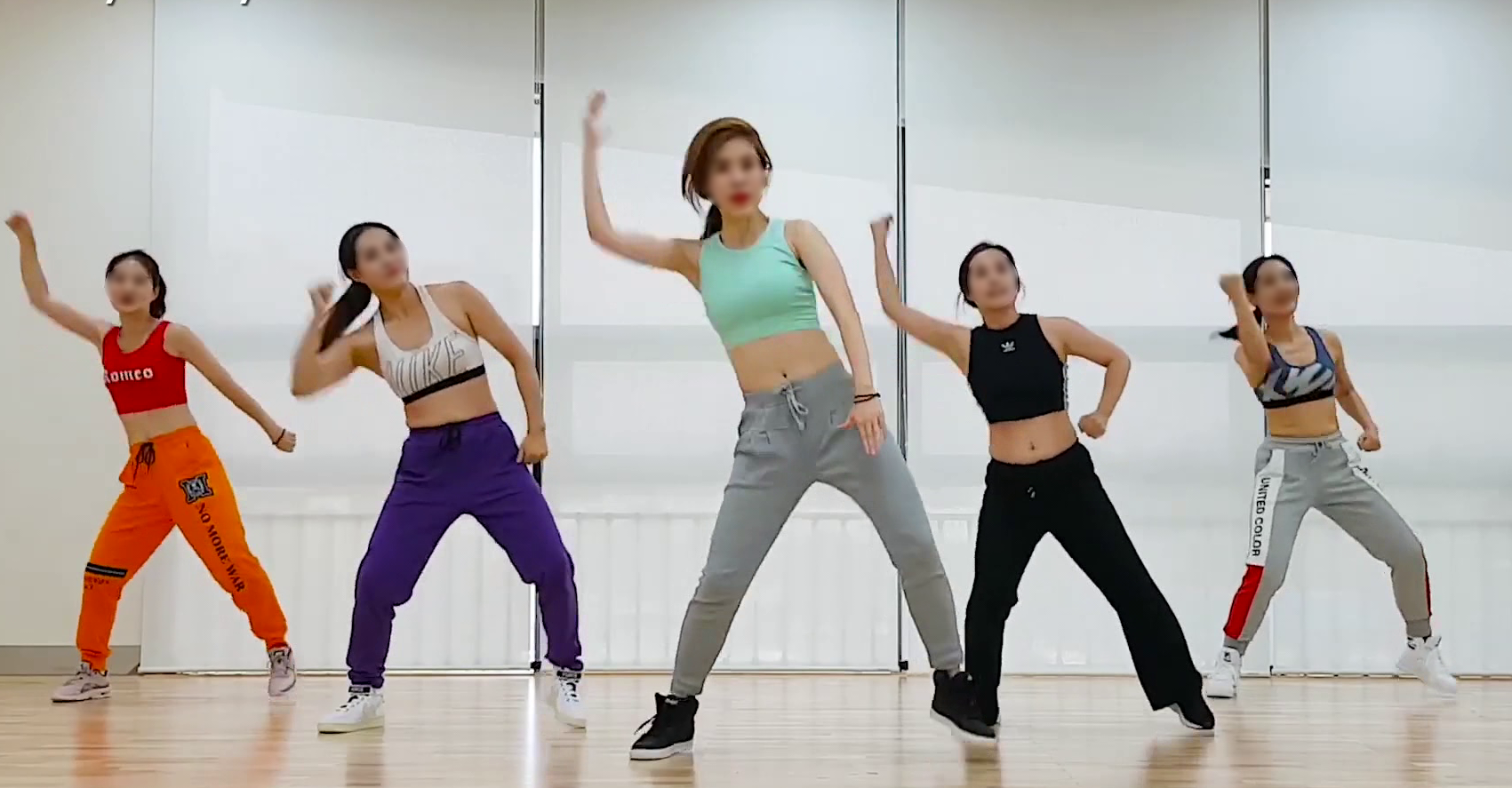}}&
\shortstack{\includegraphics[width=0.33\linewidth]{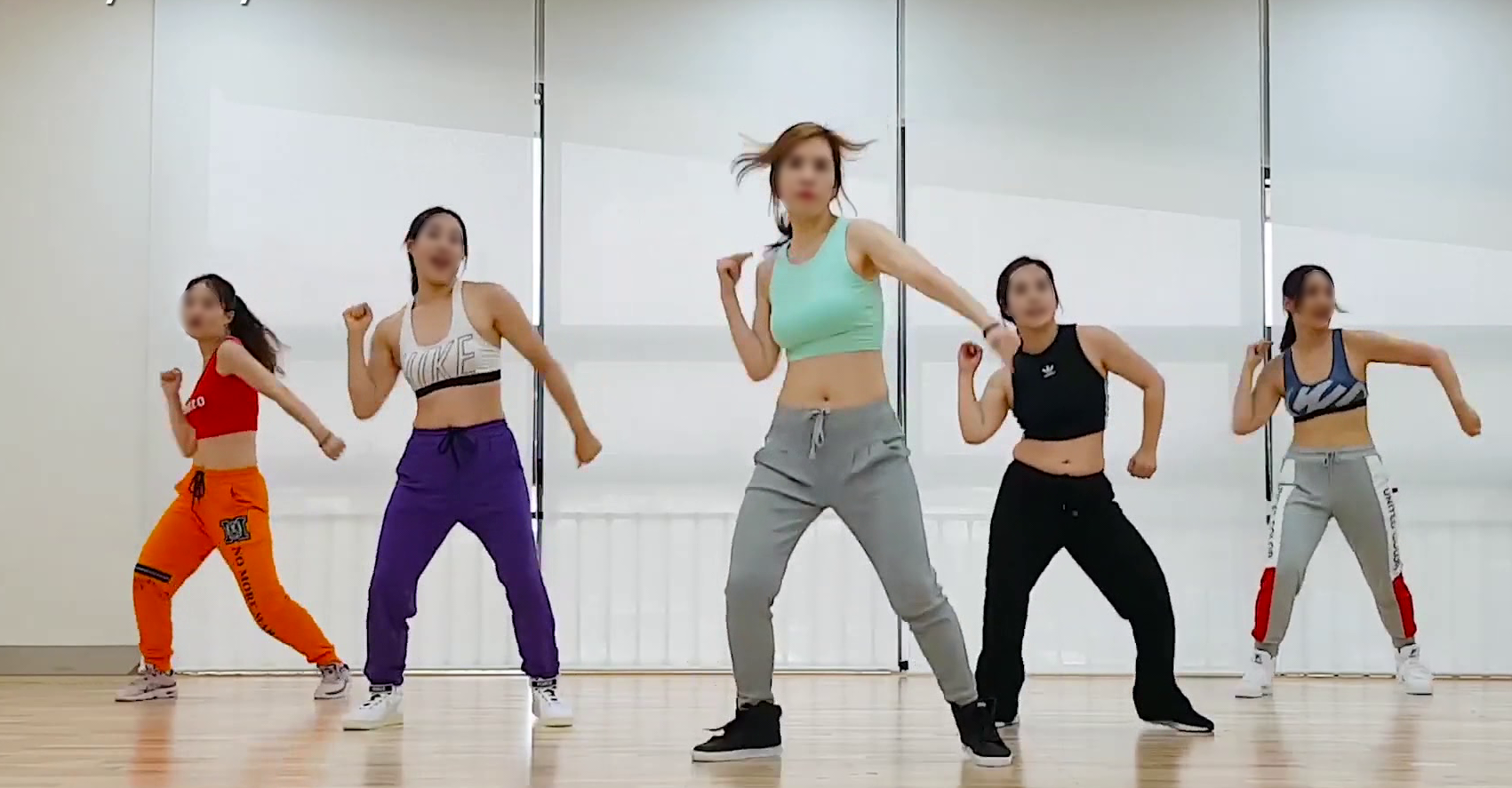}}&
\shortstack{\includegraphics[width=0.33\linewidth]{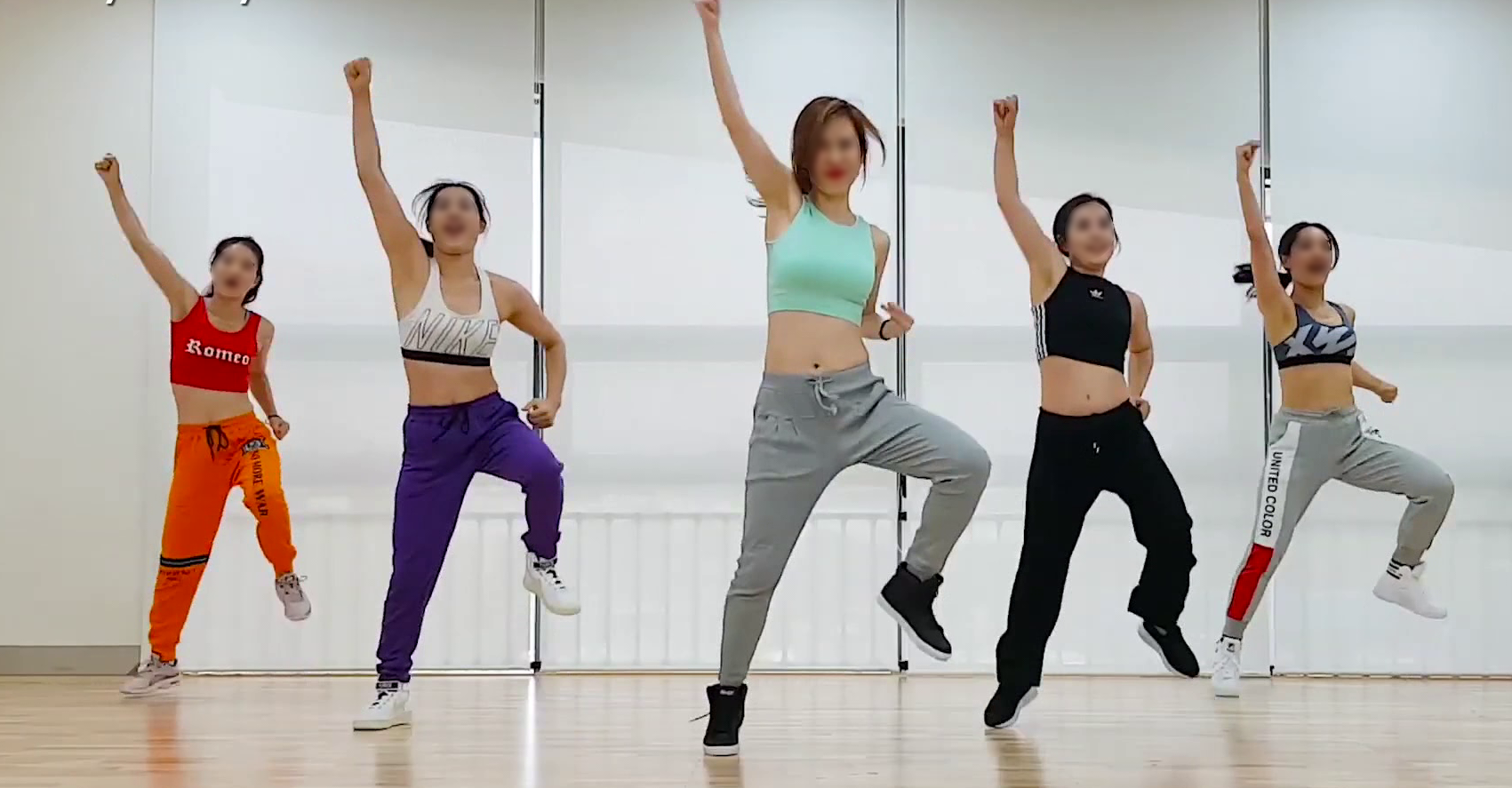}}\\[3pt]
\shortstack{\includegraphics[width=0.33\linewidth]{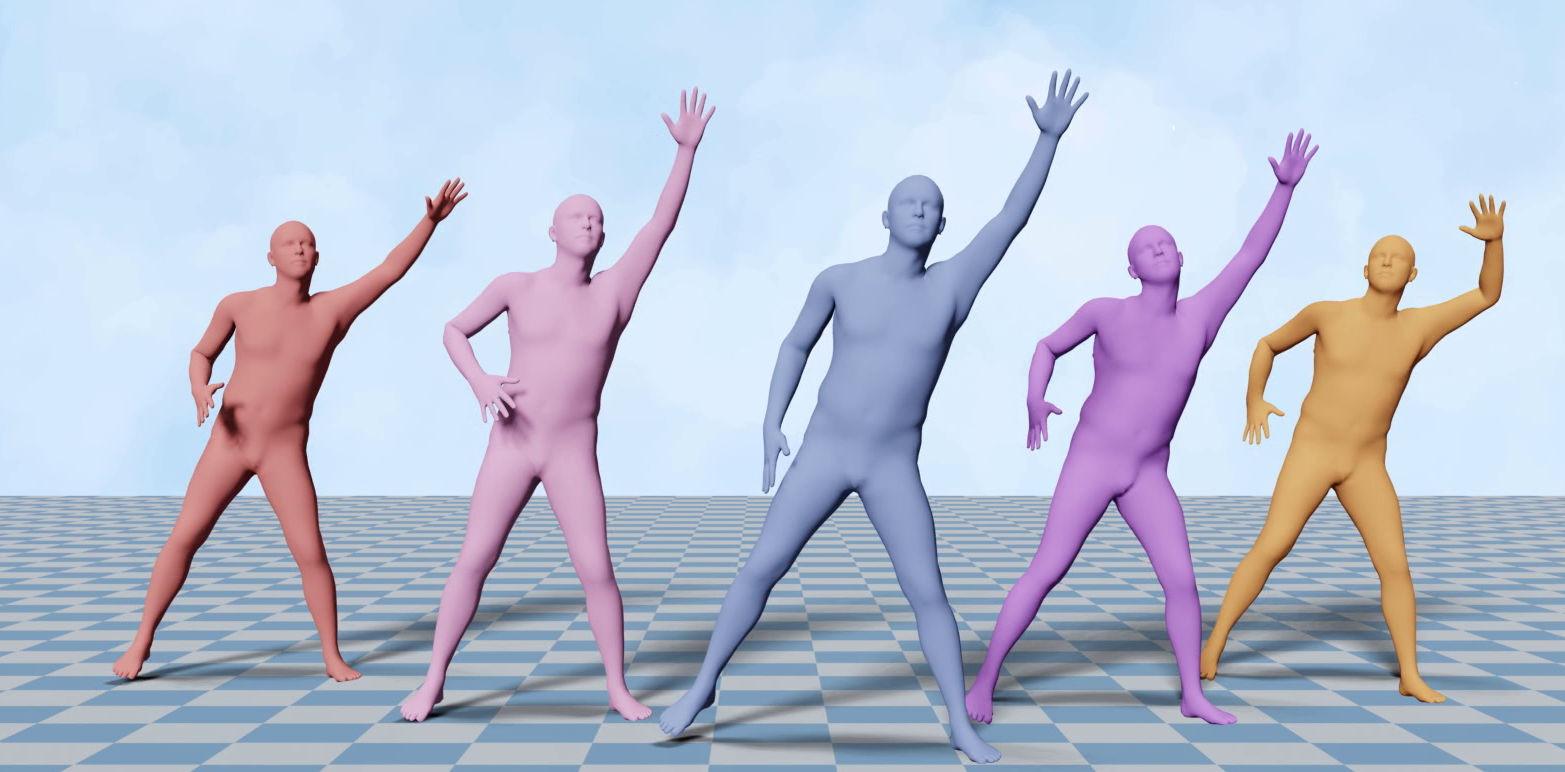}}&
\shortstack{\includegraphics[width=0.33\linewidth]{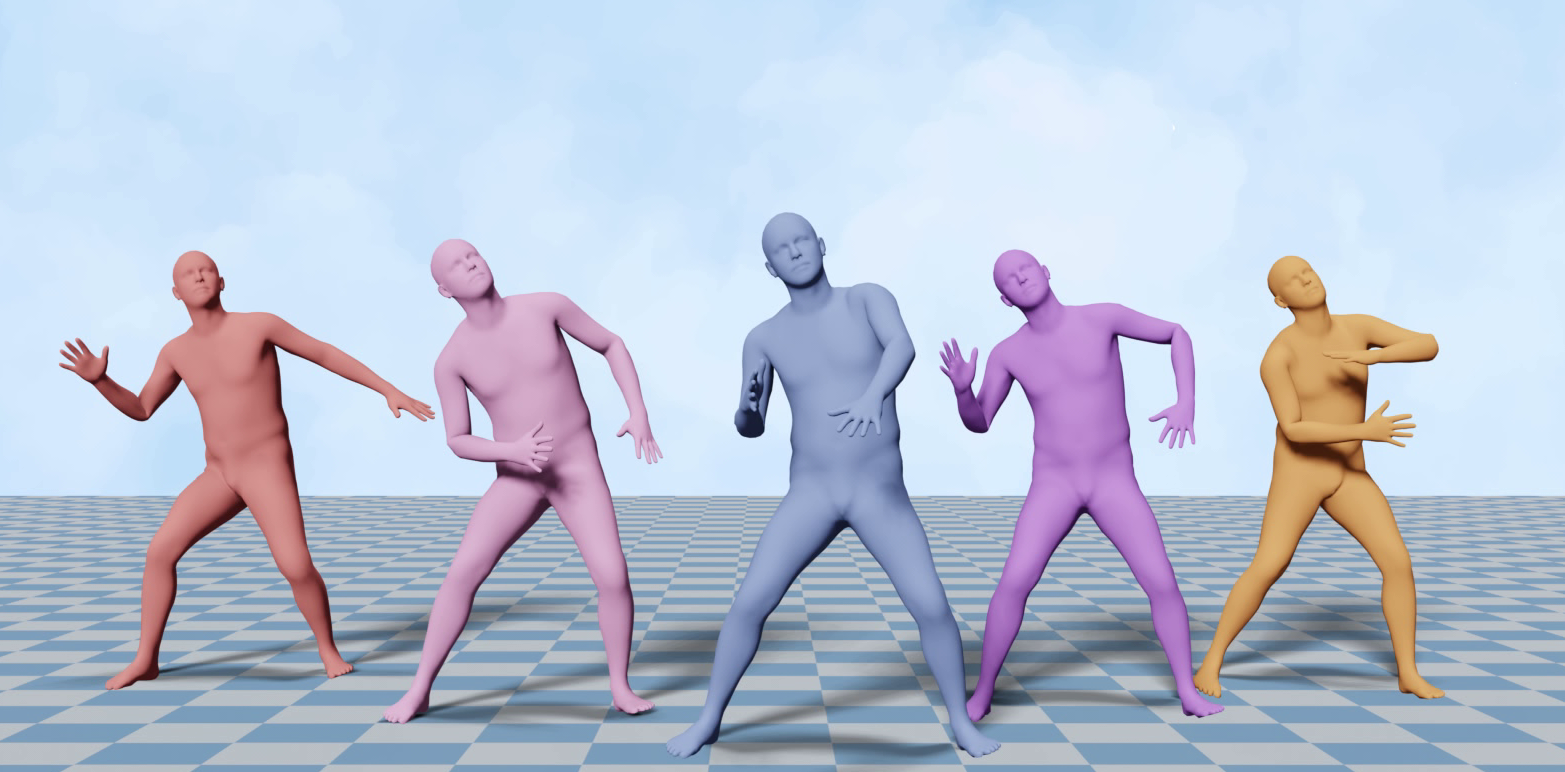}}&
\shortstack{\includegraphics[width=0.33\linewidth]{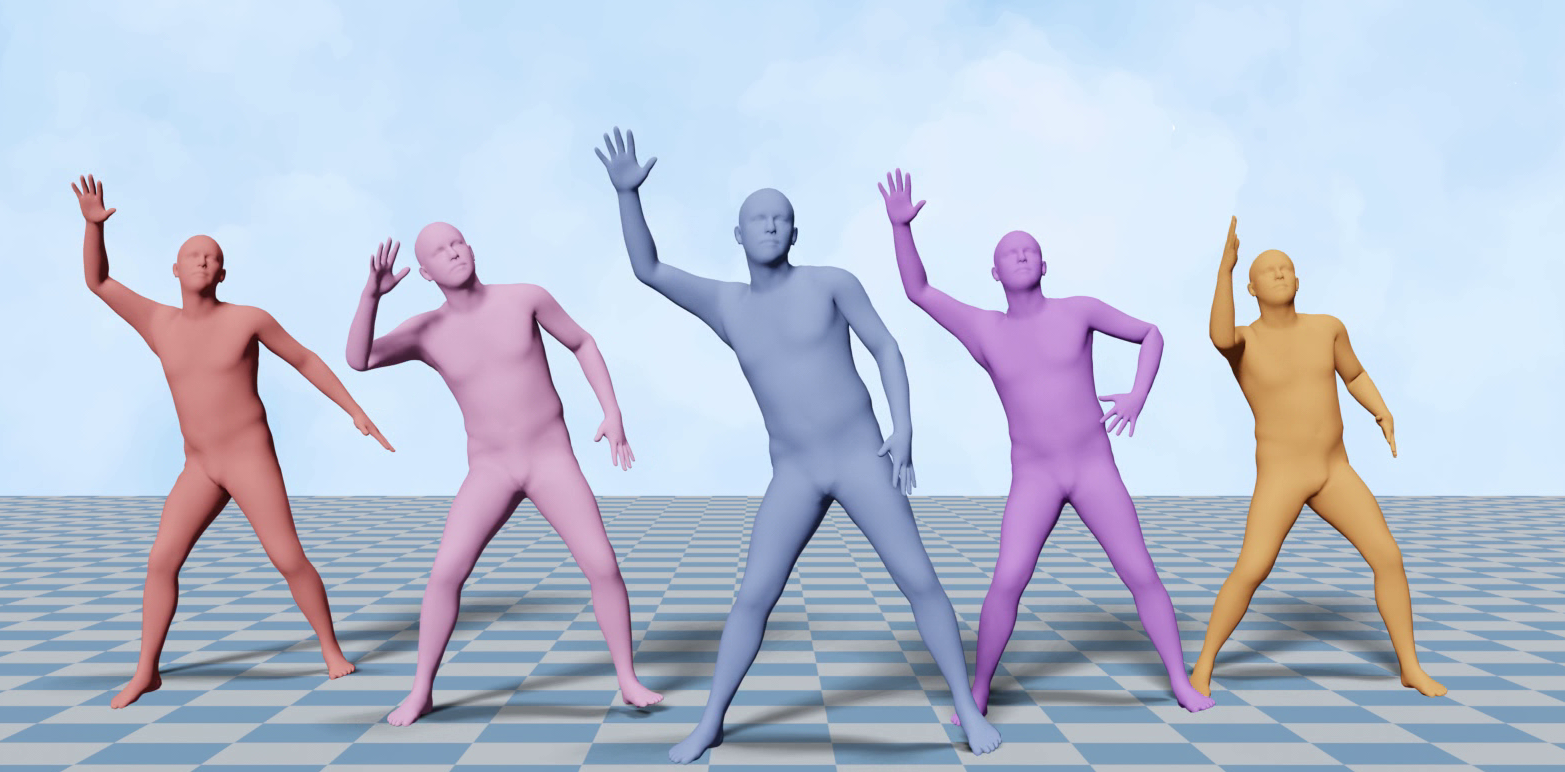}}&
\shortstack{\includegraphics[width=0.33\linewidth]{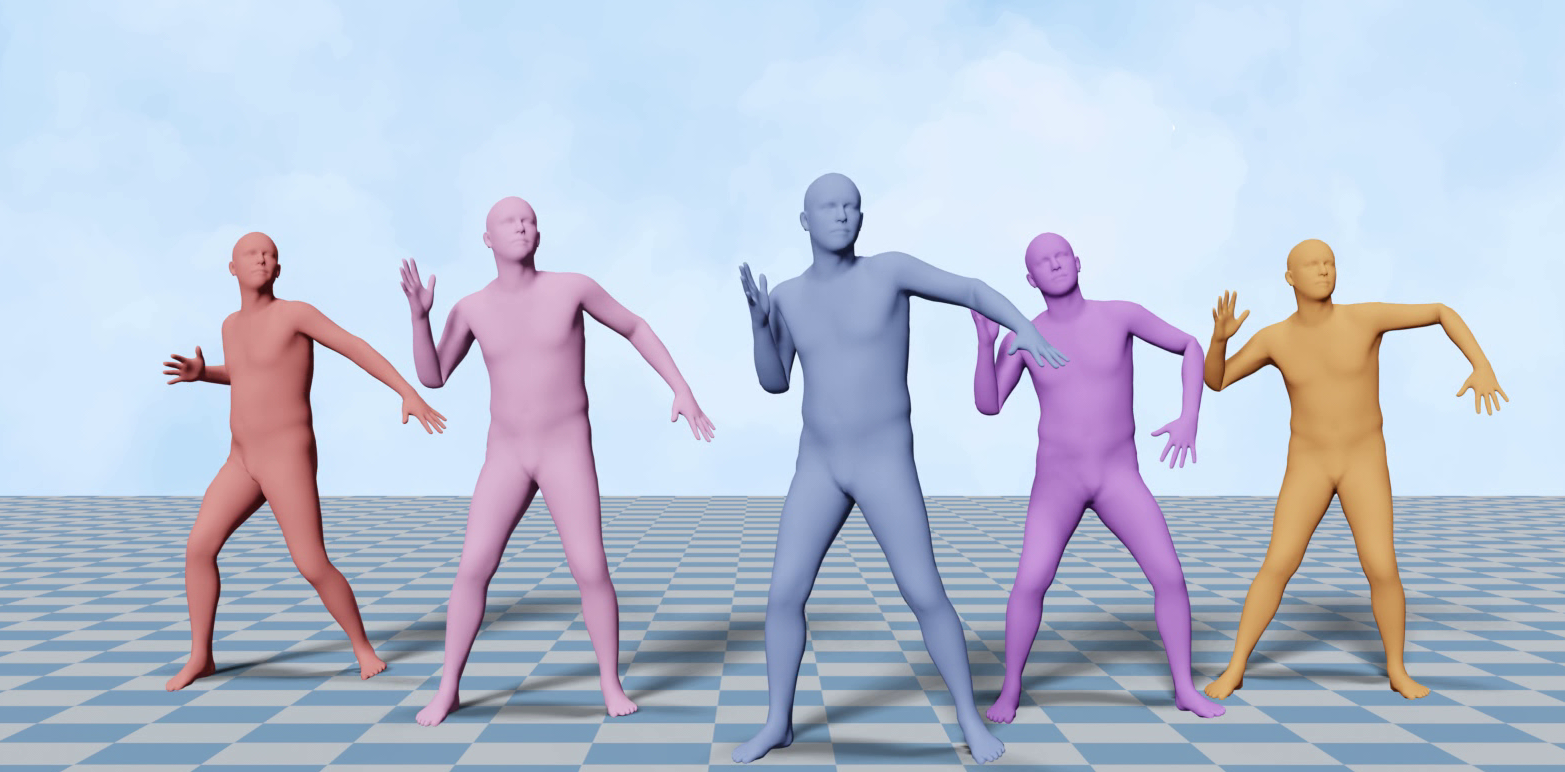}}&
\shortstack{\includegraphics[width=0.33\linewidth]{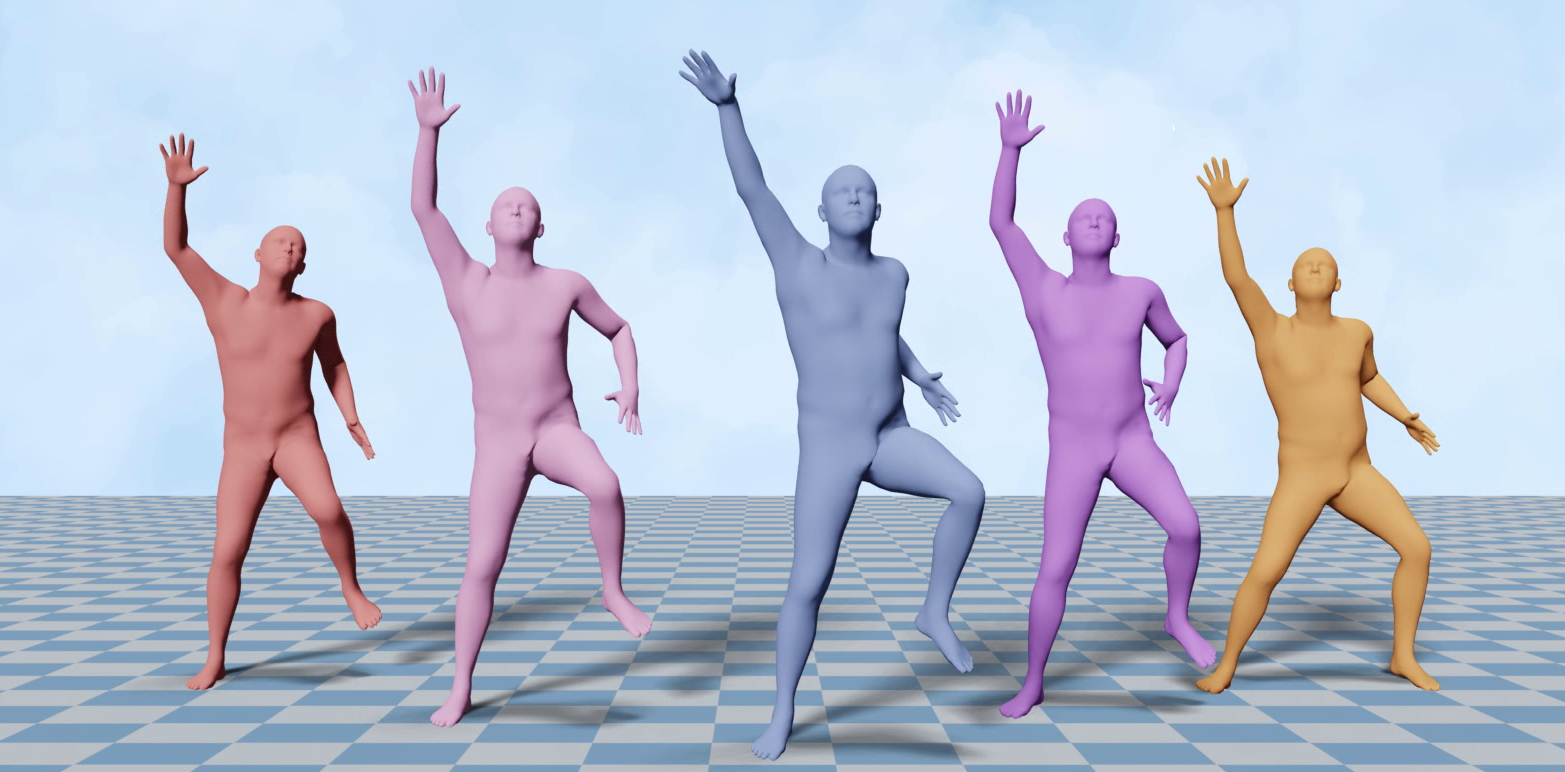}}\\[3pt]
\shortstack{\includegraphics[width=0.33\linewidth]{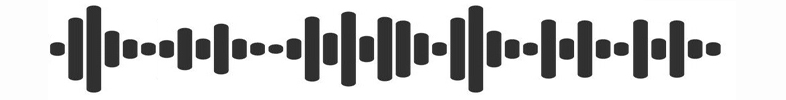}}&
\shortstack{\includegraphics[width=0.33\linewidth]{images/IntroVis/audioSignalc100.jpg}}&
\shortstack{\includegraphics[width=0.33\linewidth]{images/IntroVis/audioSignalc100.jpg}}&
\shortstack{\includegraphics[width=0.33\linewidth]{images/IntroVis/audioSignalc100.jpg}}&
\shortstack{\includegraphics[width=0.33\linewidth]{images/IntroVis/audioSignalc100.jpg}}
\end{tabular}
}
    \captionof{figure}{We present the \textsf{AIOZ-GDANCE} dataset with in-the-wild videos, music audio, and 3D group dance motion.}
    \label{fig:IntroVis}
\end{center}%
}]

\begin{abstract}
Music-driven choreography is a challenging problem with a wide variety of industrial applications. Recently, many methods have been proposed to synthesize dance motions from music for a single dancer. However, generating dance motion for a group remains an open problem. In this paper, we present $\rm AIOZ-GDANCE$, a new large-scale dataset for music-driven group dance generation. Unlike existing datasets that only support single dance, our new dataset contains group dance videos, hence supporting the study of group choreography. We propose a semi-autonomous labeling method with humans in the loop to obtain the 3D ground truth for our dataset. The proposed dataset consists of $16.7$ hours of paired music and 3D motion from in-the-wild videos, covering $7$ dance styles and $16$ music genres. We show that naively applying single dance generation technique to creating group dance motion may lead to unsatisfactory results, such as inconsistent movements and collisions between dancers. Based on our new dataset, we propose a new method that takes an input music sequence and a set of 3D positions of dancers to efficiently produce multiple group-coherent choreographies. We propose new evaluation metrics for measuring group dance quality and perform intensive experiments to demonstrate the effectiveness of our method. Our project facilitates future research on group dance generation and is available at \url{https://aioz-ai.github.io/AIOZ-GDANCE/}.
\end{abstract}

\section{Introduction}

Dancing is an important part of human culture and remains one of the most expressive physical art and communication forms~\cite{FINK2021351, Visualization-folk-dance}. With the rapid development of digital social media platforms, creating dancing videos has gained significant attention from social communities. As the consequence, millions of dancing videos are created and watched daily on online platforms. Recently, studies of how to create natural dancing motion from music have attracted great attention in the research community~\cite{bisig2022generative}. The outcome of dancing generation techniques can be applied to various applications such as animation~\cite{li2021AIST++}, virtual idol~\cite{perez2021_transflower}, metaverse~\cite{survey-dancing-deep-metaverse}, or in dance education~\cite{physic-perform, shi2021application}.

Although there is some progress towards synthesizing realistic dancing motion from music in recent literature~\cite{li2021AIST++,perez2021_transflower,siyao2022_bailando,kim2022brandnew_dance} creating natural 3D dancing motions from the input audio remains an open problem~\cite{li2021AIST++}. This is mainly due to \textit{(i)} the complex structure and the non-linear correlation between continuous human motion and the accompanying music audio, \textit{(ii)} the variety in the repertoire of dancing motions for an expressive  choreography performance, \textit{(iii)} the difficulty in generating long motion sequences, and \textit{(iv)} the complication of capturing the correspondences between the dancing motion and audio such as dancing styles or music rhythms. Furthermore, recent works focus on generating dancing motion for solo dancer~\cite{Dance_Revolution,yalta2019_weaklyrnn, li2021AIST++, ferreira2021_learn2dance_gcn} while producing dancing motion for a group of dancers have not been well-investigated by the community yet. 

Compared to the single dance generation task, the group dance generation poses a more challenging problem~\cite{eisenmann2011creating}. In practice, group dance may contain complicated and different choreographies between attended dancers while also retaining the relationship between the motion and the input music rhythmically. Furthermore, group dance has the communication between dancers through physical contact, hence performing a correlation between motion series in a group is essential and challenging. 
Currently, most of the music-to-dance datasets~\cite{li2021AIST++, perez2021_transflower, tang2018_dancemelody, zhuang2020_music2dance, sun2020_deepdance, li2022_phantomdance} only contain solo dance videos. Hence, they are not able to capture essential aspects that only occurred in group dance scenarios, including multiple-person motions, synchronization, and interaction between dancers.

Since learning to synthesize dancing motion from music is a highly challenging task, one of the primary requirements is the availability of large-scale datasets. Some works employ professional choreographers to obtain music-dance datasets under a highly complex Motion Capture (MoCap) system~\cite{chen2021_choreomaster, ye2020_choreonet, tang2018_dancemelody, li2022_phantomdance}. Although the captured motions are accurate, it is challenging to scale up and increase the diversity of the data with several dance styles and music genres.
To overcome the limitation of the MoCap system, another line of works leverage existing pose estimation algorithm to generate the pseudo-ground truths for in-the-wild dancing videos~\cite{sun2020_deepdance, Dance_Revolution,lee2019_dancing2music}. However, these aforementioned datasets are designed originally for the single motion generation task and provide only paired music and single dancing motion~\cite{li2021AIST++, li2022_phantomdance, perez2021_transflower}, thus they cannot be applied to facilitate generating multiple motions within a group of dancers.

Motivated by these shortcomings, this paper introduces \textsf{AIOZ-GDANCE}, a new large-scale dataset to advance the study of group choreography. Unlike existing choreography datasets that only supports single dancer, our dataset consists of group dance videos. As in Figure~\ref{fig:IntroVis}, our dataset has multiple input modalities  (i.e., video frames, audio) and multiple 3D human mesh ground truths. To annotate the dataset, we introduce a semi-automatic method with humans in the loop to ensure the data quality. 
Using the new dataset, we propose the first strong baseline for group dance generation that can jointly generate multiple dancing motions expressively and coherently. 

Our contributions are summarised as follows:
\begin{itemize}
\item We introduce \textsf{AIOZ-GDANCE}, a new large-scale dataset for group dance generation. To our best knowledge, \textsf{AIOZ-GDANCE} is the \textit{largest} audio-driven group dance dataset.
\item Based on our new dataset, we propose a new method, namely GDanceR, to efficiently generate group dancing motion from the input audio.
\end{itemize}

\section{Related Work}

\label{Sec:relatedwork}

\begin{table*}
\centering
\setlength{\tabcolsep}{0.3 em} 
{\renewcommand{\arraystretch}{1.0}
\begin{tabular}{l|c|c|c|c|c|c|c|c} 
\hline
{Dataset} & {Hours} & {Subjects} & \begin{tabular}[c]{@{}c@{}}{Captured}\\{Environment}\end{tabular}& {Acquisition} & \begin{tabular}[c]{@{}c@{}}{Music}\\{Genres}\end{tabular} & \begin{tabular}[c]{@{}c@{}}{Dance}\\{Styles}\end{tabular} & \begin{tabular}[c]{@{}c@{}}{Group}\\{ Dance}\end{tabular} & \begin{tabular}[c]{@{}c@{}}{Ground}\\{Truth}\end{tabular} \\ 
\hline
Dance Melody~\cite{tang2018_dancemelody} & 1.57 & n/a & Lab-control & MoCap & limited & limited & no & 3D Joints \\ 

DanceNet~\cite{zhuang2020_music2dance} & 0.96 & 2 & Lab-control & MoCap & limited & limited & no & 3D Joints \\ 

YT-Dance3D~\cite{sun2020_deepdance} & 5 & n/a & In-the-wild & Fully-automatic & rich & rich & no & 3D Joints \\ 

Dancing2Music~\cite{lee2019_dancing2music} & 71 & n/a & In-the-wild & Fully-automatic & rich & limited & no & 2D Joints \\ 

DanceRevolution~\cite{Dance_Revolution} & 12 & n/a & In-the-wild & Fully-automatic & rich & limited & no & 2D Joints \\ 

PMSD~\cite{perez2021_transflower} & 3.84 & 2 & Lab-control & MoCap &  varied  & limited & no & 3D Joints \\ 

ShaderMotionVR~\cite{perez2021_transflower} & 10.2 & 11 & In-the-wild & VR~Tracking & varied & rich & no & 3D Joints \\ 

AIST++~\cite{li2021AIST++} & 5.2 & 30 & Lab-control & Fully-automatic & limited & varied & no & 3D Mesh \\ 

PhantomDance~\cite{li2022_phantomdance} & 9.6 & 100+ & In-the-wild & Artistic & rich & rich & no & 3D Mesh \\ 
\hline \hline
\textbf{\textsf{AIOZ-GDANCE}} (ours) & \textbf{16.7} & \textbf{4000+} & \textbf{In-the-wild} & \textbf{Semi-automatic} & \textbf{rich} & \textbf{rich} & \textbf{yes} & \textbf{3D Mesh} \\
\hline
\end{tabular}}

\caption{Comparison between music-to-dance datasets.}

\label{tab:datasetCompare} 
\end{table*}

\textbf{Music-Motion Datasets.} Early efforts to create music-motion dataset focus on using motion capture system. Specifically, the authors of~\cite{tang2018_dancemelody} use MoCap to record 3D skeletons from dancers to establish a music-to-dance dataset with four dancing types. It is challenging to collect a large dataset using the MoCap method since it is costly to hire performing dancers and equip the required devices. Notice the limitations of MoCap, the authors in~\cite{lee2019_dancing2music,Dance_Revolution} propose a music-to-dance dataset by crawling internet videos and use OpenPose~\cite{openpose1} to obtain 2D skeleton ground truths. The authors in~\cite{perez2021_transflower} propose a dataset using MoCap and Virtual Reality data. Instead of generating dance ground truths as human skeletons, AIST++ dataset~\cite{li2021AIST++,sun2019deepAISTorigin} obtains 3D motion by fitting SMPL model~\cite{SMPL:2015} from multi-view videos. Unlike~\cite{li2021AIST++} the work in~\cite{li2022_phantomdance} proposes a large-scale music-to-dance dataset from in-the-wild videos. Nevertheless, existing datasets only focus on the single-dance scenario. Therefore, the group motion and the interaction between dancers are not exploited in these datasets. We compare different music-to-dance datasets in Table~\ref{tab:datasetCompare}.

\textbf{Audio-driven Motion Generation.}
Generating natural and realistic human motion from audio is a challenging problem~\cite{joshi2021extensive}. A classical approach is based on the motion graph constructed from a largely captured motion database~\cite{motion_graph1}. To synthesize novel motion, different motion segments are combined by optimizing the transition cost along the path of the graph~\cite{motion_graph2}. Other works apply music-motion similarity matching constraints to further ensure the consistency between motion and music \cite{motion_graph3,motion_graph4,motion_graph5,motion_graph6,moion_graph7,thangstyle}. In recent years, several progresses have been made in the field of music-to-dance motion generation using CNN~\cite{zhuang2020_music2dance, sun2020_deepdance,ye2020_choreonet,ahn2020_autoregressive}, RNN~\cite{tang2018_dancemelody, sun2020_deepdance, alemi2017_groovenet, Dance_Revolution}, GCN~\cite{ferreira2021_learn2dance_gcn, ren2020_ssl_gcn}, GAN~\cite{sun2020_deepdance,lee2019_dancing2music}, or Transformer~\cite{siyao2022_bailando, li2021AIST++, li2022_phantomdance, perez2021_transflower,kim2022brandnew_dance}. Typically, these music-to-dance methods are conditioned on multimodal inputs 
and then generate the future sequence of human poses.
Despite the potential to generate realistic dancing motion from the music, these methods lack the ability to produce coherent movements between multiple dancers altogether. 

\textbf{Multi-person Motion Prediction.} Learning and forecasting the behavior between multiple people has been a longstanding problem~\cite{stergiou2019analyzing,khaire2022deep,nguyen2021graph}. Alahi~\etal~\cite{alahi2014socially} jointly reasons the trajectories of multiple pedestrians and forecasts their destinations in a scene by using Markov chain model. The authors in~\cite{adeli2020socially} combine the visual context of the scene and social interactions between multiple persons to forecast their future motion. Recently, the Multi-Range Transformers proposed by~\cite{wang2021multi} can predict the movements of more than 10 people for social interaction groups. 
Difference from the motion prediction task that uses motion or visual inputs, we aim to generate human dancing motion from the input music.

\section{The \textsf{AIOZ-GDANCE} Dataset}

Since we want to develop a large-scale dataset with in-the-wild videos, setting up a MoCap system is not feasible. However, manually annotating 3D groundtruth for millions of frames from dancing videos is also an extremely tedious job. Therefore, we propose a semi-automatic labeling method with humans in the loop to produce a large-scale group dance dataset.

\subsection{Data Collection and Preprocessing}
\label{sec:tracking}
\textbf{Video Collection.} We collect the in-the-wild, public domain group dancing videos along with the music from Youtube, Tiktok, and Facebook.   
All group dance videos are processed at $1920 \times 1080$ resolution and 30FPS.

\textbf{Human Tracking.} We perform tracking for all humans in the videos using the state-of-the-art multi-object tracker~\cite{sun2022_dancetrack} to obtain the tracking bounding boxes. Note that although the tracker can produce reasonable results, there are failure cases in some frames. Therefore, we manually correct the bounding box of the incorrect cases. This tracking correction is crucial since we want the trajectory of each person to be accurately tracked in order to reconstruct their motion in latter stages.

\textbf{Pose Estimation.} Given the bounding boxes of each person in the video, we leverage the recent 2D pose estimation method~\cite{alphapose1} to generate the initial 2D poses for each person. 
In practice, there exist some inaccurately detected keypoints due to motion blur and partial occlusion. We manually fix the incorrect cases to obtain the 2D keypoints of each human bounding box.

\begin{figure*}[ht]
    \centering
    \includegraphics[width=\textwidth,keepaspectratio=true]{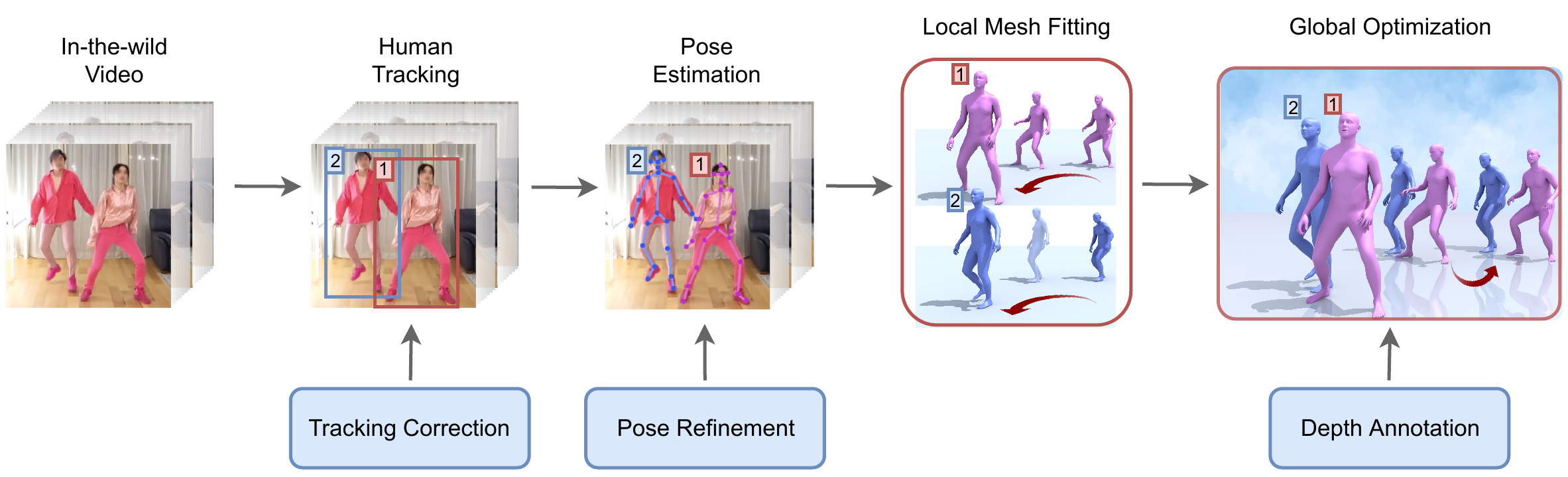}
    \vspace{-3ex}
    \caption{The pipeline of making our \textsf{AIOZ-GDANCE} dataset. Blue boxes denote manual correction/annotation steps.}
    \vspace{-1ex}
    \label{fig:overview_dataset_pipeline}
\end{figure*}

\subsection{Group Motion Fitting}
To construct 3D group dance motion, we first reconstruct the full body motion for each dancer by fitting the 3D mesh. We then jointly optimize all dancer motions to construct the globally-coherent group motion. Finally, we post-process and remove wrong cases from the optimization results.

\textbf{Local Mesh Fitting.}
\label{sec:mesh-fitting} 
We use SMPL model~\cite{SMPL:2015} to represent the 3D human. The SMPL model is a differentiable function that maps the pose parameters $\mathbf{\theta}$, the shape parameters $\mathbf{\beta}$, and the root translation $\mathbf{\tau}$ into a set of 3D human body mesh vertices $\mathbf{V}\in \mathds{R}^{6890\times3}$ and 3D joints $\mathbf{X}\in \mathds{R}^{J\times3}$, where $J$ is the number of body joints.
Our optimizing motion variables for each individual dancer consist of a sequence of SMPL joint angles $\{\mathbf{\theta}_t\}_{t=1}^T$, a sequence of the root translation $\{\mathbf{\tau}_t\}_{t=1}^T$, and a single SMPL shape parameter $\mathbf{\beta}$. We fit the sequence of SMPL motion variables to the tracked 2D keypoints (obtained from Section \ref{sec:tracking}) by extending SMPLify-X~\cite{SMPL-X:2019} across the whole video sequence:
\begin{equation}
\label{eq:mesh_fitting}
E_{\rm local} = E_{\rm J} + \lambda_{\theta}E_{\theta} + \lambda_{\beta} E_{\beta} + \lambda_{\rm S}E_{\rm S} + \lambda_{\rm F}E_{\rm F},
\end{equation}
where $E_{\rm J}$, $E_{\theta}$ and $E_{\beta}$ are as in~\cite{SMPL-X:2019} but calculated across every frames of the video sequence. The smoothness term $E_{\rm S} = \sum_{t=1}^{T-1}\Vert \mathbf{\theta}_{t+1} - \mathbf{\theta}_{t} \Vert^2 + \sum_{j=1}^J\sum_{t=1}^{T-1}\Vert \mathbf{X}_{j,t+1} - \mathbf{X}_{j,t} \Vert^2$ encourages the temporal smoothness of the motion. The term $E_{\rm F} =  \sum_{t=1}^{T-1} \sum_{j \in \mathcal{F}} c_{j,t}\Vert \mathbf{X}_{j,t+1} - \mathbf{X}_{j,t} \Vert^2$ ensures feet joints to stay stationary when in contact (zero velocity). $\mathcal{F}$ is the set of feet joint indexes, $c_{j,t}$ is the feet contact label of joint $j$ at time $t$ produced by a contact estimation network~\cite{zou2020contact_net}.

\textbf{Global Optimization.}
Given the 3D motion sequence of each dancer $p$: $\{\mathbf{\theta}^p_t, \mathbf{\tau}^p_t\}$, we further resolve the motion trajectory problems in group dance by solving the following objective: 
\begin{align}
\label{eq:global_opt}
E_{\rm global} &= E_{\rm J} + \lambda_{\rm pen}E_{\rm pen} + \lambda_{\rm reg}\sum_{p}E_{\rm reg}(p) \notag\\ &+ \lambda_{\rm dep}\sum_{p,p',t}E_{\rm dep}(p,p',t) + \lambda_{\rm gc}\sum_{p}E_{\rm gc}(p),
\end{align}
where 
$E_{\rm pen}$ is the Signed Distance Function penetration term based on~\cite{jiang2020_coherent_reconstruction} to prevent the overlapping of reconstructed motions between dancers.
${E_{\rm reg}(p) =\sum_{t=1}^T\Vert \mathbf{\theta}^p_t - \hat{\mathbf{\theta}}^p_t\Vert^2}$ is the regularization term that prevents the motion from deviating too much from the prior optimized individual motion $\{\hat{\mathbf{\theta}}^p_t\}$ obtained by optimizing Equation \ref{eq:mesh_fitting} for dancer $p$.

In practice, we find that the relative depth ordering of dancers in the scene can be inconsistent due to the ambiguity of the 2D projection. To ensure the group motion quality, we watch the videos and manually provide the ordinal depth relation information of all dancers in the scene at each frame $t$ as follows:
\begin{equation}
r_t(p,p') =
\begin{cases}
1, &\text{if dancer } p \text{ is closer than } p' \\ 
-1, &\text{if dancer } p \text{ is farther than } p' \\
0, &\text{if their depths are roughly equal}
\end{cases}
\end{equation}

Given the relative depth information provided by human annotators, we derive the depth relation term $E_{\rm dep}$ inspired by~\cite{chen2016_depth_ranking}. This term encourages consistent ordinal depth relation between the motion trajectories of multiple dancers, especially when dancers partially occlude each other:  
\begin{equation}
\small
\label{eq:global_depth1}
E_{\rm dep}(p,p',t) =
\begin{cases}
     \log(1+\exp(z^p_t - z^{p'}_t)), &r_t(p,p')=1 \\
     \log(1+\exp(-z^p_t + z^{p'}_t)), &r_t(p,p')=-1 \\
     (z^p_t - z^{p'}_t)^2, &r_t(p,p')=0 \\
\end{cases}
\end{equation}
where $z^p_t$ is the depth component of the root translation $\mathbf{\tau}^p_t$ of the person $p$ at frame $t$.

Finally, we apply the global ground contact constraint $E_{\rm gc}$ to further ensure consistency between the motion of every person and the environment based on the ground contact information. This contact term is also needed to reduce the artifacts such as foot-skating, jittering, and penetration under the ground.
\begin{equation}
\label{eq_Egc}
\small
E_{\rm gc}(p) = \sum_{t=1}^{T-1} \sum_{j \in \mathcal{F}} c^p_{j,t}\Vert \mathbf{X}^p_{j,t+1} - \mathbf{X}^p_{j,t} \Vert^2 + c^p_{j,t} \Vert  (\mathbf{X}^p_{j,t} - \mathbf{f})^\top \mathbf{n}^* \Vert^2,
\end{equation}
where $\mathcal{F}$ is the set of feet joint indexes, $\mathbf{n}^*$ is the estimated plane normal and $\mathbf{f}$ is a 3D fixed point on the ground plane. The first term in Equation~\ref{eq_Egc} is the zero velocity constraint when the feet are in contact with the ground, while the second term encourages the feet position to stay near the ground when in contact. To obtain the ground plane parameters, we initialize the plane point $\mathbf{f}$ as the weighted median of all contact feet positions. The plane normal $\mathbf{n}^*$ is obtained by optimizing a robust Huber objective:
\begin{equation}
\small
\mathbf{n}^* = \arg\min_{\mathbf{n}} \sum_{\mathbf{X}_{\rm feet}} \mathcal{H}\left((\mathbf{X}_{\rm feet} - \mathbf{f})^\top \frac{\mathbf{n}}{\Vert\mathbf{n}\Vert}\right) + \Vert \mathbf{n}^\top\mathbf{n} - 1 \Vert^2,
\end{equation}
where $\mathcal{H}$ is the Huber loss function~\cite{huber1992robust},  $\mathbf{X}_{\rm feet}$ is the 3D feet positions of all dancers across the whole sequence that are labelled as in contact (i.e., $c^p_{j,t} = 1$) .

\textbf{Post Processing.}
Although our optimization process produces relatively good results, there are some extreme cases that it fails to handle. We recheck all the results and fix the cases with minor problems. Other severely wrong cases are simply discarded. More details can be found in our Supplementary Material.

\begin{figure*}[ht]
    \centering
    \includegraphics[width=\textwidth]{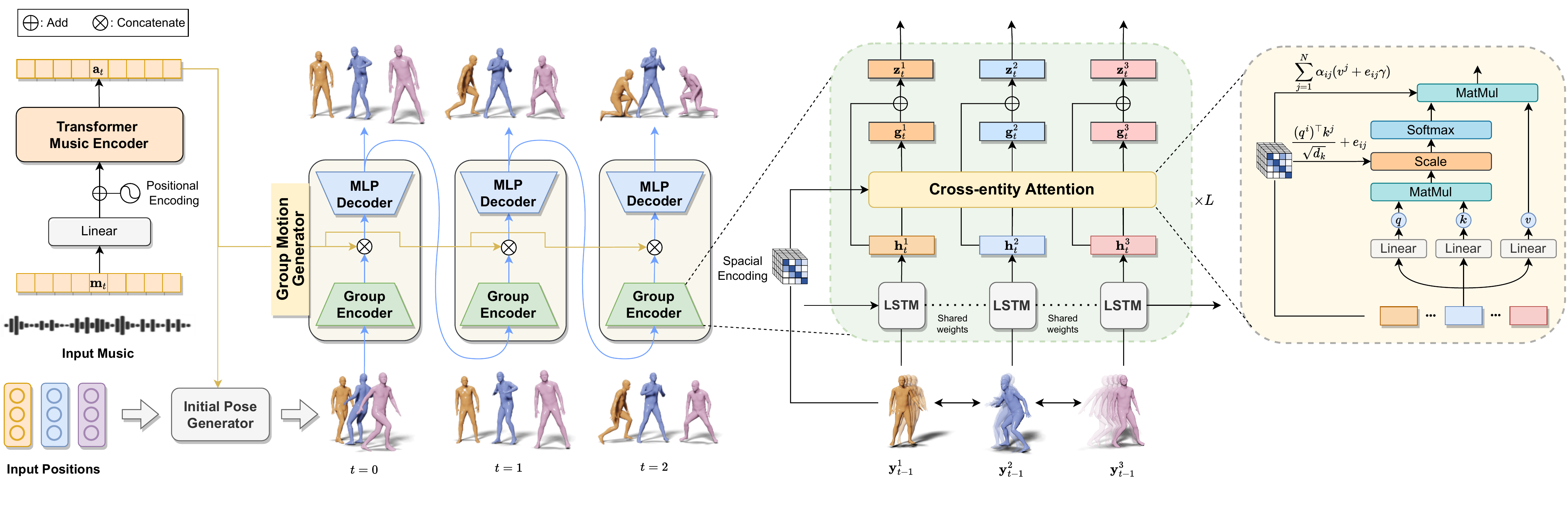}
    \caption{
    Architecture of our Music-driven 3D \textbf{G}roup \textbf{Dance} generato\textbf{r} (GDanceR). Our model takes in a music sequence and a set of initial positions, and then auto-regressively generates coherent group dance motions that are attuned to the input music.
    }
    \label{fig:GMG}
\end{figure*}

\subsection{How will \textbf{\textsf{AIOZ-GDANCE}} be useful to the community?}

We bring up some interesting research directions that can be benefited from our dataset:

\begin{itemize}
    \item Group Dance Generation: While single-person choreography is a hot research topic~\cite{survey-dancing-deep-metaverse,li2021AIST++,siyao2022_bailando,li2022_phantomdance,chen2021_choreomaster}, group dance generation has not yet well investigated. We hope that the release of our dataset will foster more this research direction.

    \item Human Pose Tracking: By having SMPL groundtruth motion, our dataset can be used in many human pose/motion tracking tasks such as in~\cite{yang2021learning,sun2022_dancetrack,doering2022posetrack21}. 
    
\end{itemize}

Apart from these tasks, we believe our dataset can be used in other scenarios such as dance education~\cite{papillon2022pirounet,ferreira2021_learn2dance_gcn}, dance style transfer~\cite{zhang2021dance,yin2022dance,thangstyle}, or human behavior analysis~\cite{men2022gan,le2022global,lee2022human,le2023uncertainty}. The research community is free to explore other applications of our dataset.

\section{Audio-driven Group Dance Generation}

\subsection{Problem Formulation}
\label{sec:task}
Given an input music audio sequence $\{m_1, m_2, ...,m_T\}$ with $t = \{1,..., T\}$ indicates the index of music segments, and the initial 3D positions of $N$ dancers $\{\tau^1_0, \tau^2_0, ..., \tau^N_0 \}$, $\tau^i_0 \in \mathds{R}^{3}$, our goal is to generate the group motion sequences $\{y^1_1,..., y^1_T; ...;y^n_1,...,y^n_T\}$ where $y^i_t$ is the generated pose of $i$-th dancer at time step $t$. Specifically, we represent the human pose as a 72-dimensional vector $y = [\tau;  \theta]$ where $\tau$, $\theta$ represent the root translation and pose parameters of the SMPL model~\cite{SMPL:2015}, respectively.

Generally, the generated group dance motion should meet the two conditions: \textit{(i)} consistency between the generated dancing motion and the input music in terms of style, rhythm, and beat; \textit{(ii)} the motions and trajectories of dancers should be coherent without cross-body intersection between dancers. Figure~\ref{fig:GMG} shows an overview of our approach.

\subsection{GDanceR Architecture}

\textbf{Transformer Music Encoder.} From the raw audio signal of the input music, we first extract music features using the audio processing library~\cite{mcfee2015librosa}. Concretely, we extract the mel frequency cepstral coefficients (MFCC), MFCC delta, constant-Q chromagram, tempogram, onset strength and one-hot beat, which results in a 438-dimensional feature vector. We then encode the music sequence $\{m_1, m_2, ...,m_T\}$, $m_t \in \mathds{R}^{438}$ into a sequence of hidden representation $\{a_1, a_2,..., a_T\}$, $a_t \in \mathds{R}^{d_a}$. In practice, we utilize the self-attention mechanism of transformer~\cite{vaswani2017attention} to effectively encode the multi-scale information and the long-term dependency between music frames. The hidden audio at each time step is expected to contain meaningful structural information to ensure that the generated dancing motion is coherent across the whole sequence.

\textbf{Initial Pose Generator.}
Given the initial positions of all dancers, we generate the initial poses by combing the audio feature with the starting positions. 
We aggregate the audio representation by taking an average over the audio sequence.
The aggregated audio is then concatenated with the input position and fed to a multilayer perceptron (MLP) to predict the initial pose for each dancer:
\begin{equation}
\label{eq:initial_pose_generator}
y^i_0 = \text{MLP}\left( \left[\frac{1}{T}\sum_{t=1}^T a_t ; \tau^i_0 \right] \right),
\end{equation}
where $[;]$ is the concatenation operator, $\tau^i_0$ is the initial position of the $i$-th dancer.

\textbf{Group Motion Generator.}
\label{sub_sec_group_motion_generator} 
To generate the group dance motion, we aim to synthesize the coherent motion of each dancer such that it aligns well with the input music. Furthermore, we also need to maintain global consistency between all dancers. In practice, our Group Encoder uses Recurrent Neural Network to capture the temporal motion dynamics of each dancer~\cite{li2018_rnn_motion_prediction, mao2019_rnn_motion_prediction}, and attention mechanism to encode the spatial relationship of all dancers~\cite{vaswani2017attention}. 

Specifically, at each time step, the pose of each dancer in the previous frame $y^i_{t-1}$ is sent to an LSTM unit~\cite{lstm} to encode the hidden local motion representation, i.e., ${h^i_t=\text{LSTM}(y^i_{t-1},h^i_{t-1})}$. 

To ensure the motions of all dancers have global coherency without strange effects such as cross-body intersection, we introduce the Cross-entity Attention mechanism. In particular, each individual motion representation is first linearly projected into a key vector $k^i$, a query vector $q^i$ and a value vector $v^i$ as follows: 
\begin{equation}
    k^i = h^i W^{k}, \quad q^i = h^i W^{q}, \quad v^i = h^i W^{v},
\end{equation}
where $W^q, W^k \in \mathds{R}^{d_h \times d_k}$, and  $W^v \in \mathds{R}^{d_h \times d_v}$ are parameters that transform the hidden motion $h$ into a query, a key, and a value, respectively. $d_k$ is the dimension of the query and key while $d_v$ is the dimension of the value vector. To encode the relationship between dancers in the scene, our Cross-entity Attention also utilizes the Scaled Dot-Product Attention as in \cite{vaswani2017attention}.

In practice, we find that people having closer positions to each other tend to have higher correlation in their movement. Therefore, we use Spacial Encoding to encode the spacial relationship between two dancers. The Spacial Encoding between two entities based on their distance in the 3D space is defined as follows:
\begin{equation}
\label{eq:e_ij}
    e_{ij} = \exp\left(-\frac{\Vert \tau^i - \tau^j \Vert^2}{\sqrt{d_{\tau}}}\right),
\end{equation}
where $d_{\tau}$ is the dimension of the position vector $\tau$. Considering the query $q^i$, which represents the current entity information, and the key $k^j$, which represents other entity information, we inject the spatial relation information between these two entities onto their cross attention coefficient: 
\begin{align}
    \alpha_{ij} &= \text{softmax}\left(\frac{(q^i)^\top k^j}{\sqrt{d_k}} + e_{ij}\right).
\end{align}

To preserve the spatial relative information in the attentive representation, we also embed them into the hidden value vector and obtain the global-aware representation $g^i$ of the $i$-th entity as follows:
\begin{align}
    g^i &= \sum_{j=1}^N\alpha_{ij}(v^j +  e_{ij}\gamma),
\end{align}
where $\gamma \in \mathds{R}^{d_v}$ is the learnable bias and scaled by the Spacial Encoding. Intuitively, the Spacial Encoding acts as the bias in the attention weight, encouraging the interactivity and awareness to be higher between closer entities. Our attention mechanism can adaptively attend to each dancer and others temporally and spatially, thanks to the encoded motion as well as the spatial information.

We then fuse both the local and global motion representation by adding $h^i$ and $g^i$ to obtain the final latent motion $z^i$. Our final global-local representation of each entity is expected to carry the comprehensive information of their own past motion as well as the motion of every other entity, enabling the MLP Decoder to generate coherent group dancing sequences. Finally, we generate the next movement ${y}^i_t$ based on the final motion representation $z^i_t$ as well as the hidden audio representation $a_t$, and thus can capture the fine-grained correspondence between music feature sequence and dance movement sequence:
\begin{equation}
\label{eq:decoder}
y^i_t = \text{MLP}([z^i_t; a_t]).
\end{equation}

\section{Experiments}
\subsection{\textbf{\textsf{AIOZ-GDANCE}} Statistic}
\label{sec:dataAnalysis}

\begin{table}[!t]
\centering
\resizebox{\linewidth}{!}{
\setlength{\tabcolsep}{0.2 em}
{\renewcommand{\arraystretch}{1.2}
\begin{tabular}{l|c|c|c|c}
\hline
\multicolumn{1}{l|}{{Criteria}} & {Train} & {Validate} & {Test} & {Total} \\ \hline
Duration (hours) &$13.5$  &$1.6$  & $1.6$ &$16.7$   \\ \hline
Total Frames & $1,459$K & $175$K  & $174$K  & $1,808$K \\ \hline
\end{tabular}
}
}
    \caption{Train/val/ test split of our \textsf{AIOZ-GDANCE} dataset.}
    \label{tab:datasetSplit}
\end{table}

\begin{figure}[!t]
 \centering	
\subfloat[]{
	\begin{minipage}[c]{
	   0.23\textwidth}
	   \label{fig_stat_music_genre}
	   \centering
	   \includegraphics[width=0.97\textwidth]{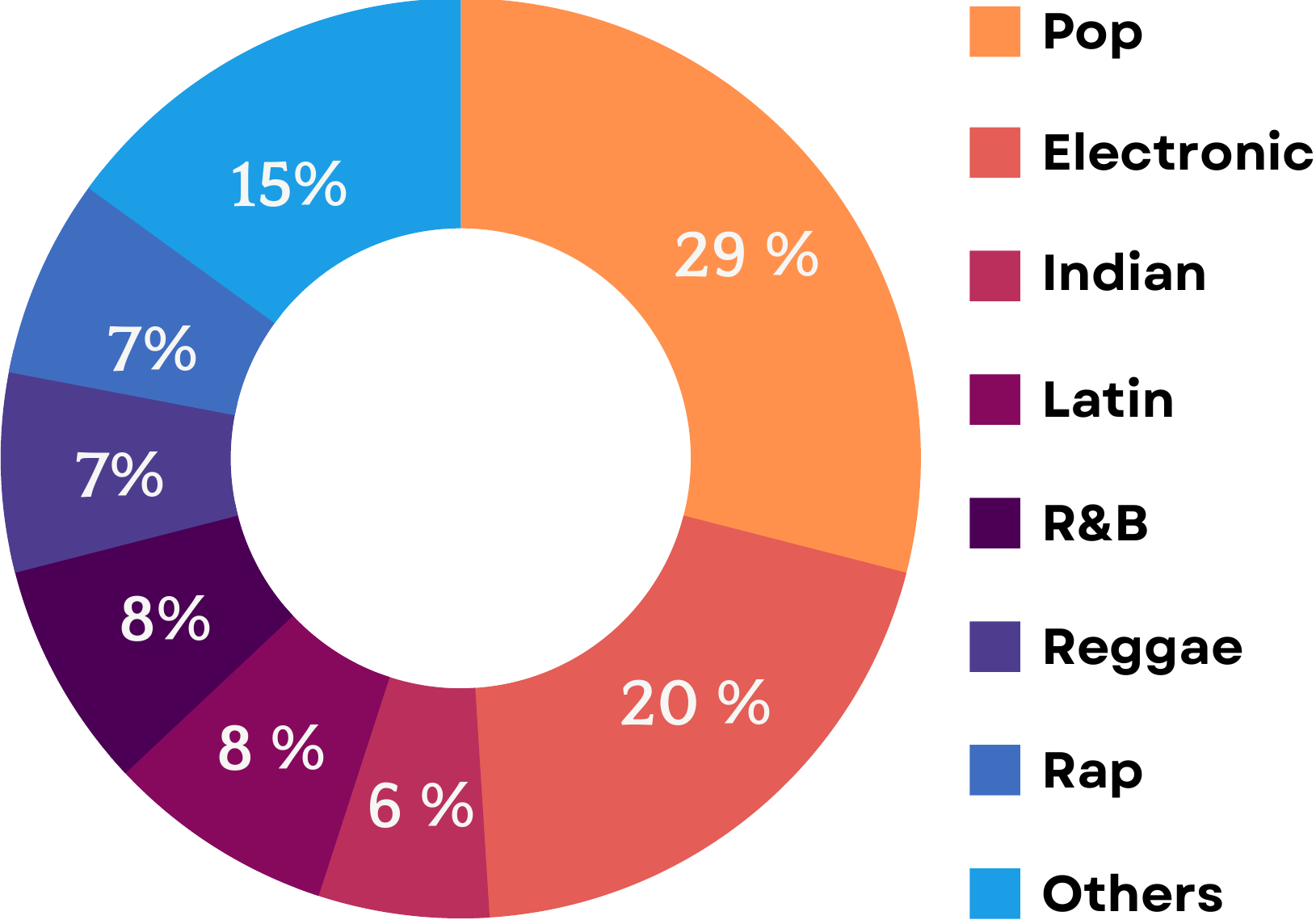}
	\end{minipage}}
 \hspace{0ex}	
  \subfloat[]{
	\begin{minipage}[c]{
	   0.23\textwidth}
	   \label{fig_stat_dance_style}
	   \centering
	   \includegraphics[width=1.0\textwidth]{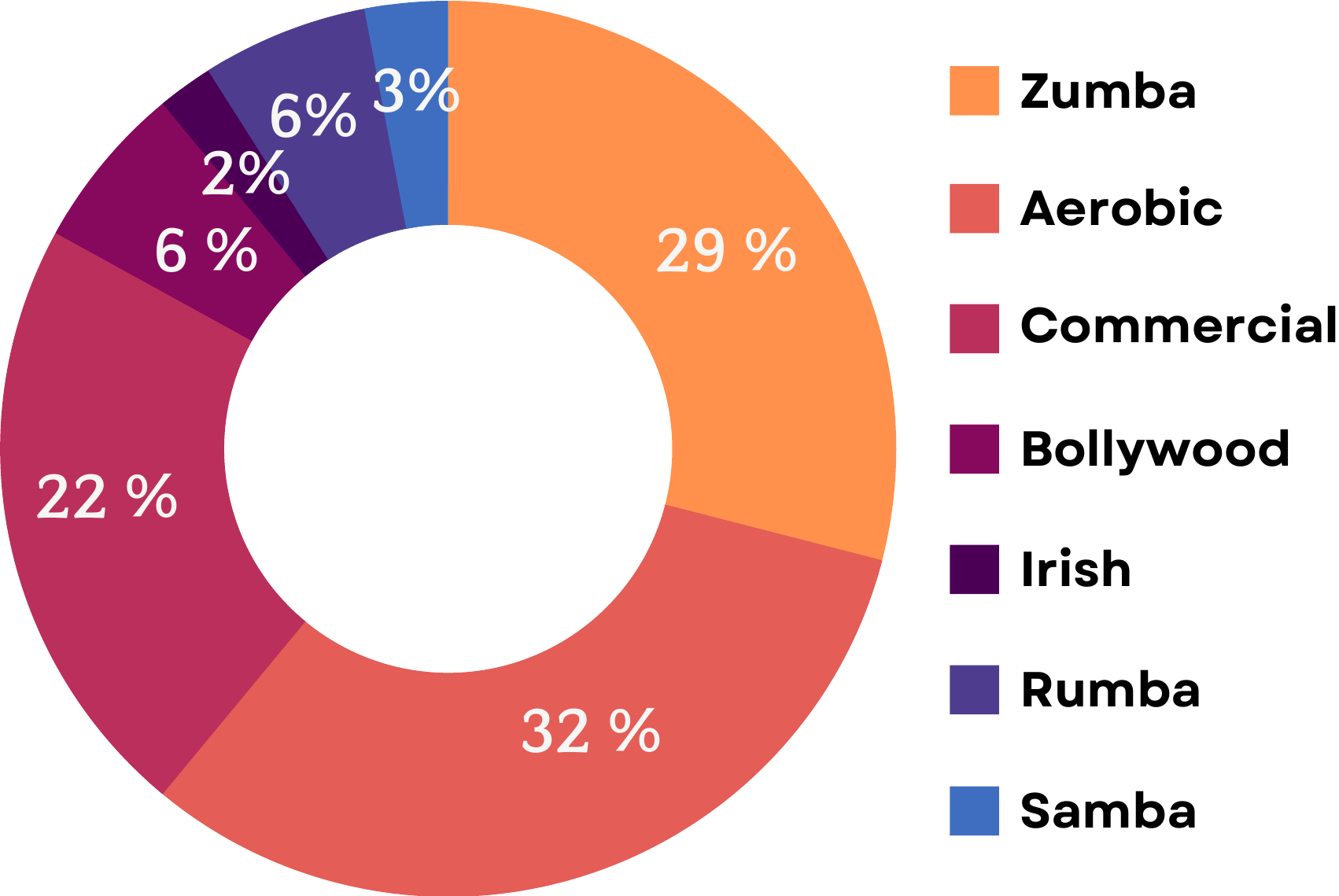}
	\end{minipage}}
\caption{Distribution (\%) of music genres (a) and dance styles (b) in our dataset.}
\label{fig:vis_Full_Balanced_Stat}
\end{figure}

\begin{figure}[!t]
 \centering	
\subfloat[]{
	\begin{minipage}[c]{
	   0.23\textwidth}
	   \label{fig_corr_dancegroup_number}
	   \centering
	   \includegraphics[width=1\textwidth]{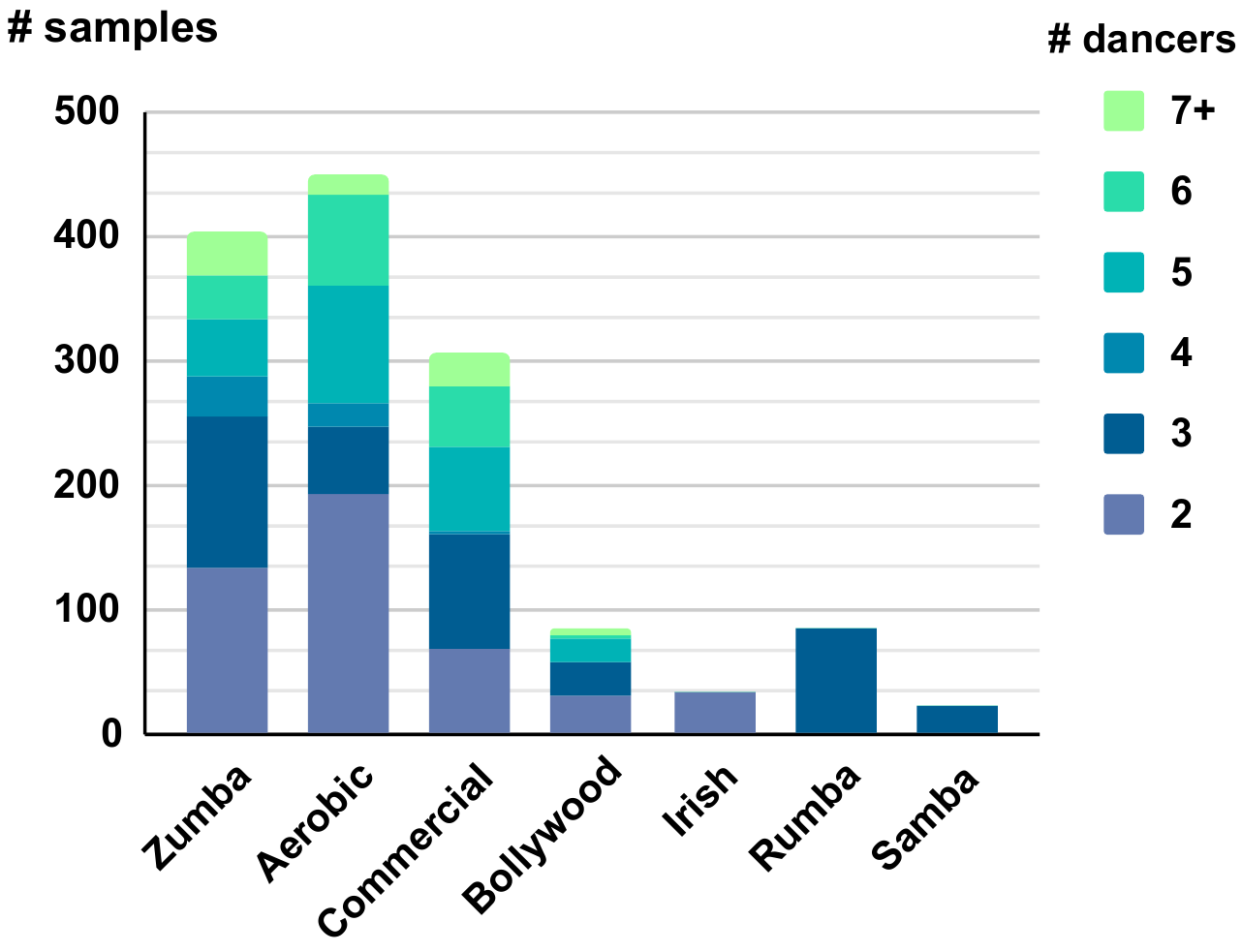}
	\end{minipage}}
 \hspace{0ex}	
  \subfloat[]{
	\begin{minipage}[c]{
	   0.23\textwidth}
	   \label{fig_corr_dancemusic}
	   \centering
	   \includegraphics[width=1.0\textwidth]{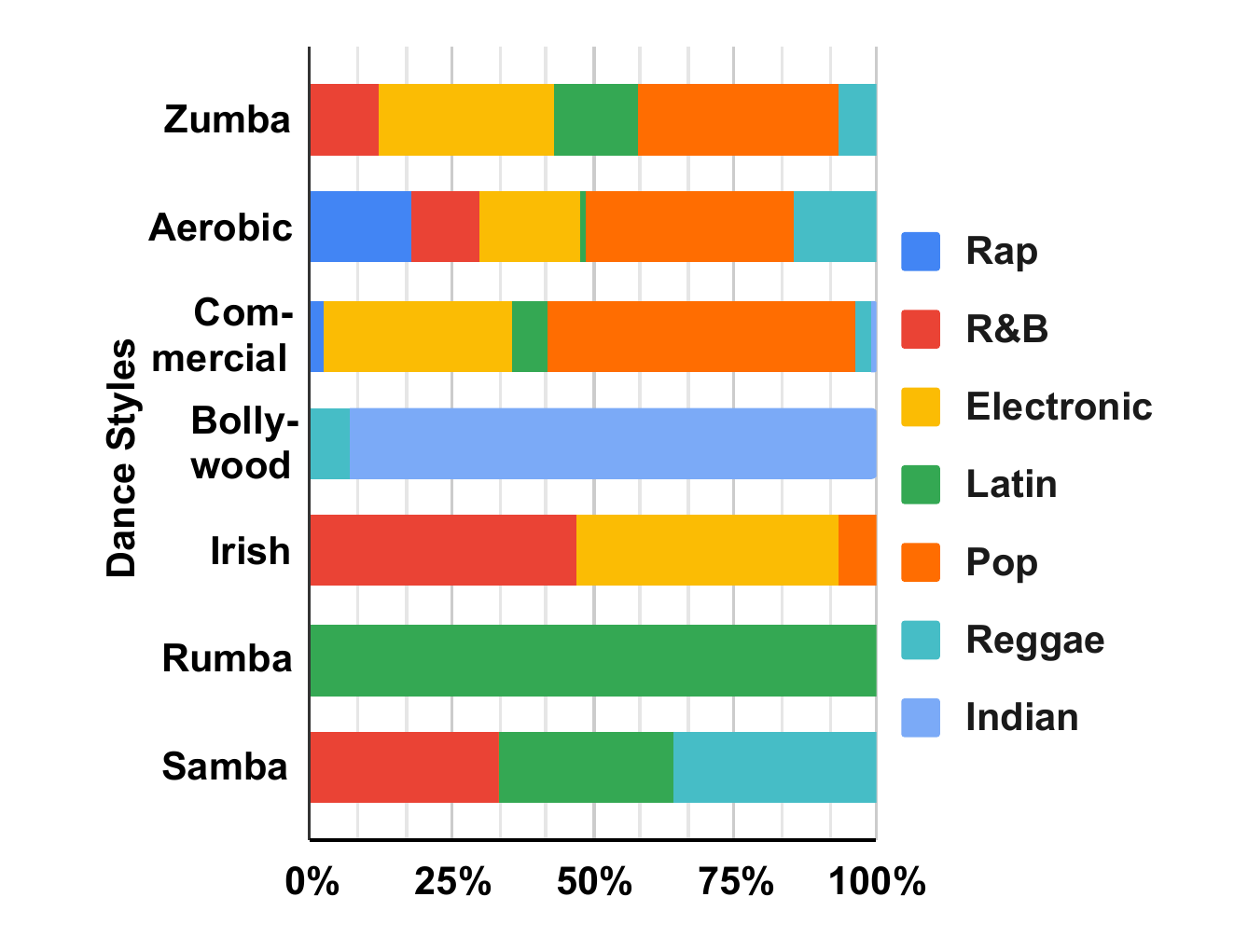}
	\end{minipage}}
\caption{The correlation between dance styles and number of dancers (a); and between dance styles and music genres (b).}
\label{fig_correlation}
\end{figure}

\textbf{Dataset Split.} \textsf{AIOZ-GDANCE} comprises $16.7$ hours of whole-body motion and music audio of group dancing.
The duration of each video in our dataset is ranging from $15$ to $60$ seconds. We decode all videos at $30$ FPS.
We randomly sample all videos into train, validation and test sets with $80\%$; $10\%$; and $10\%$ of total videos, respectively. Table~\ref{tab:datasetSplit} shows the details about the training, validation, and testing splits of our dataset.

\begin{figure*}[ht] 
  \centering
  \Large
\resizebox{\linewidth}{!}{
\setlength{\tabcolsep}{2pt}
\begin{tabular}{ccccccc}
\shortstack{\includegraphics[width=0.33\linewidth]{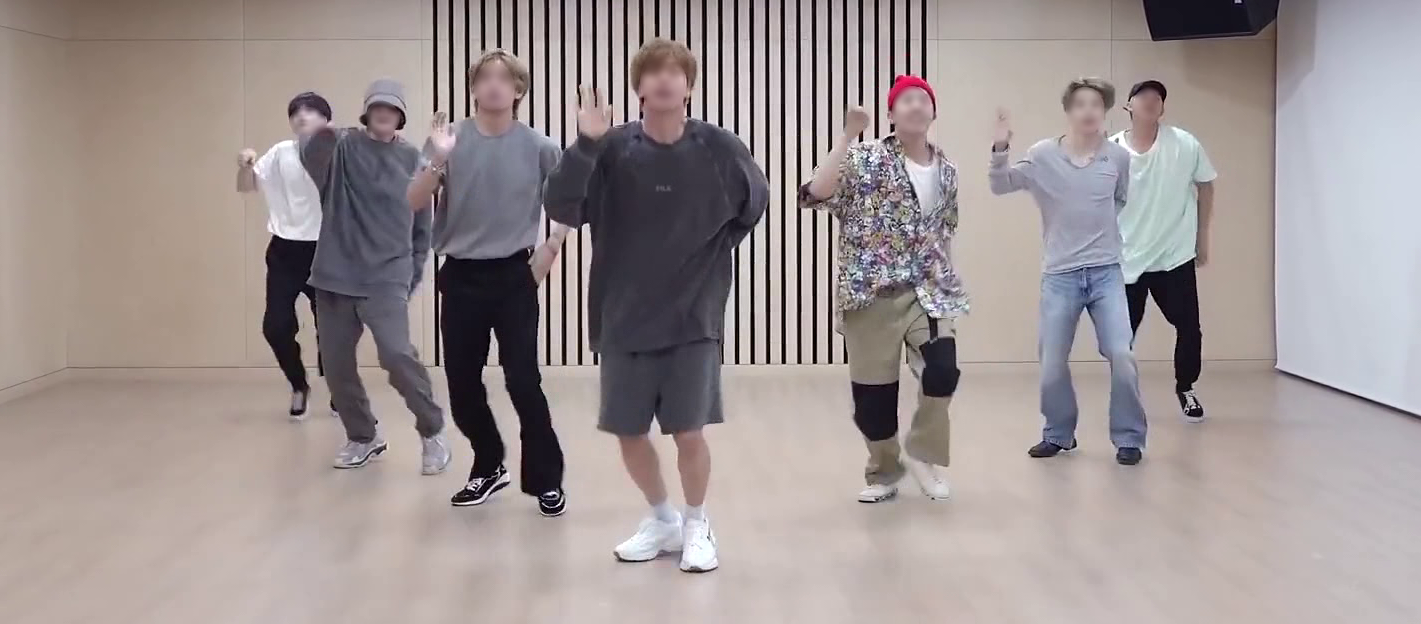}}&
\shortstack{\includegraphics[width=0.33\linewidth]{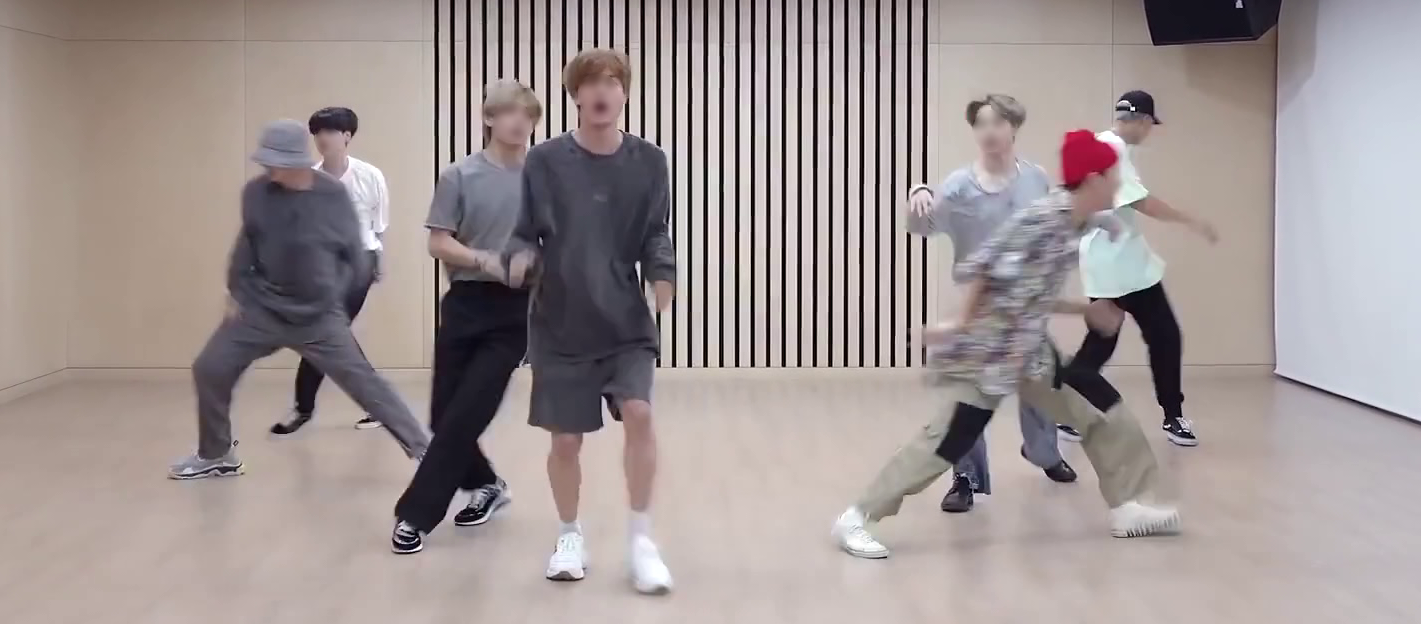}}&
\shortstack{\includegraphics[width=0.33\linewidth]{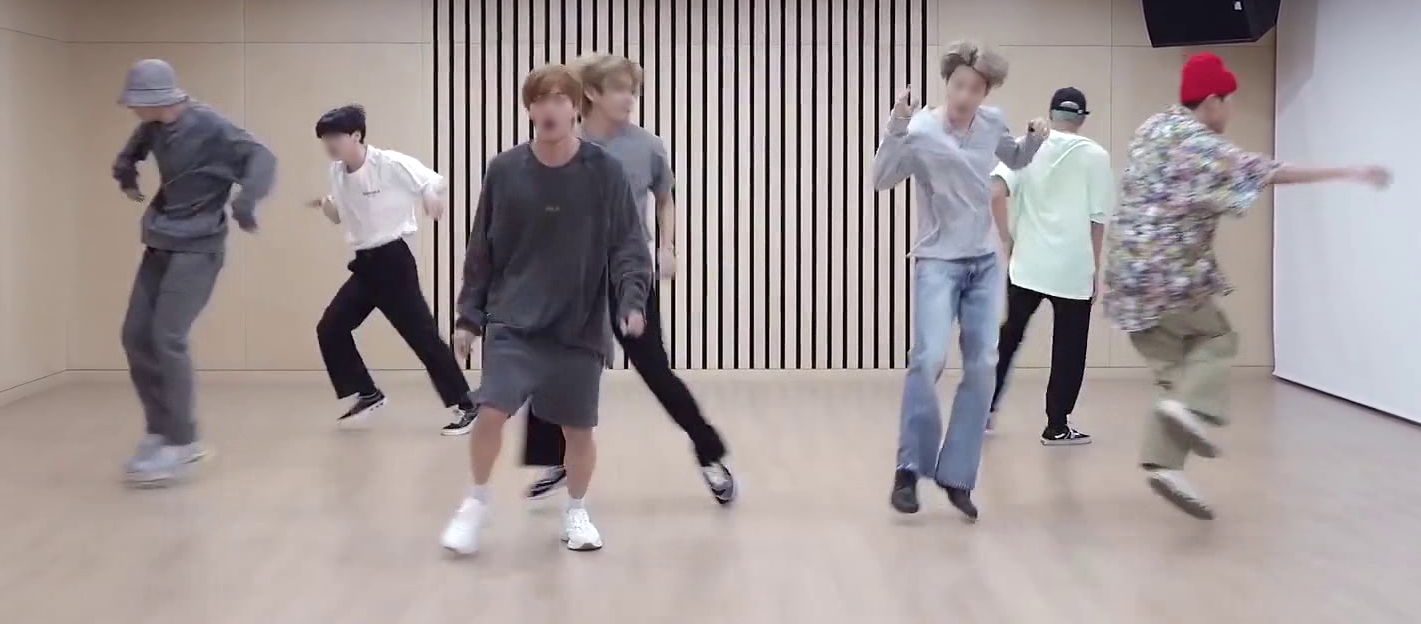}}&
\shortstack{\includegraphics[width=0.33\linewidth]{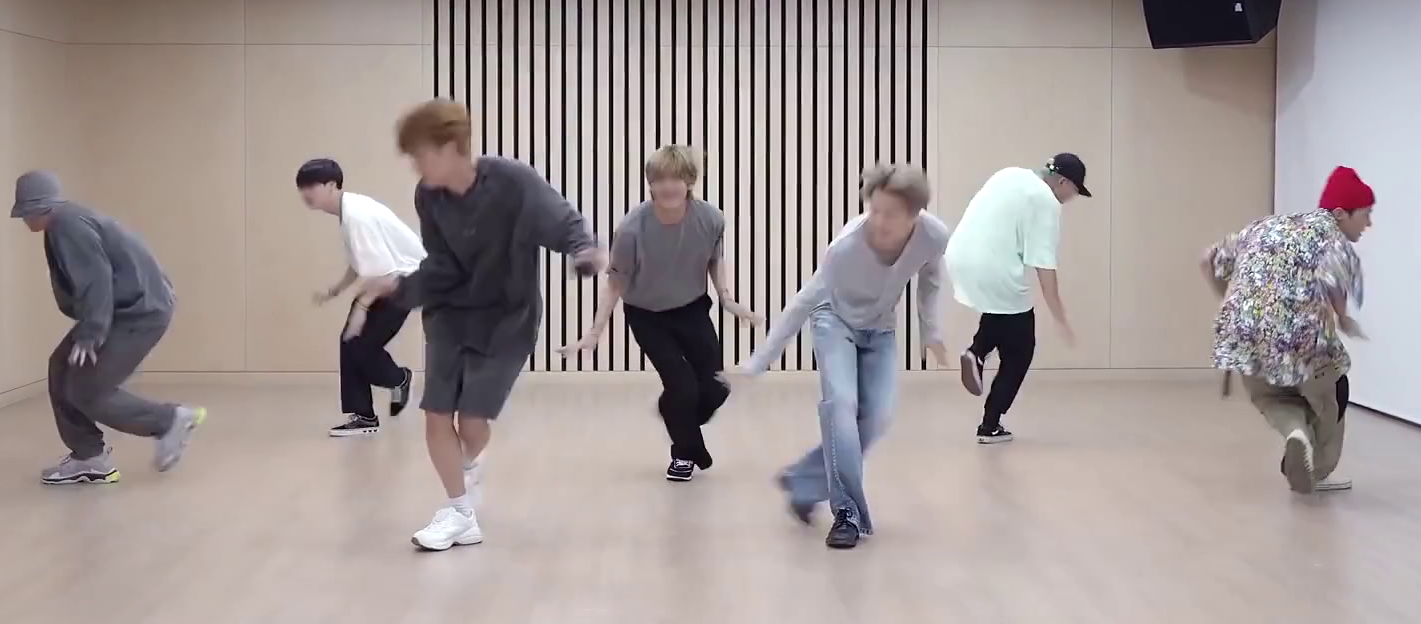}}&
\shortstack{\includegraphics[width=0.33\linewidth]{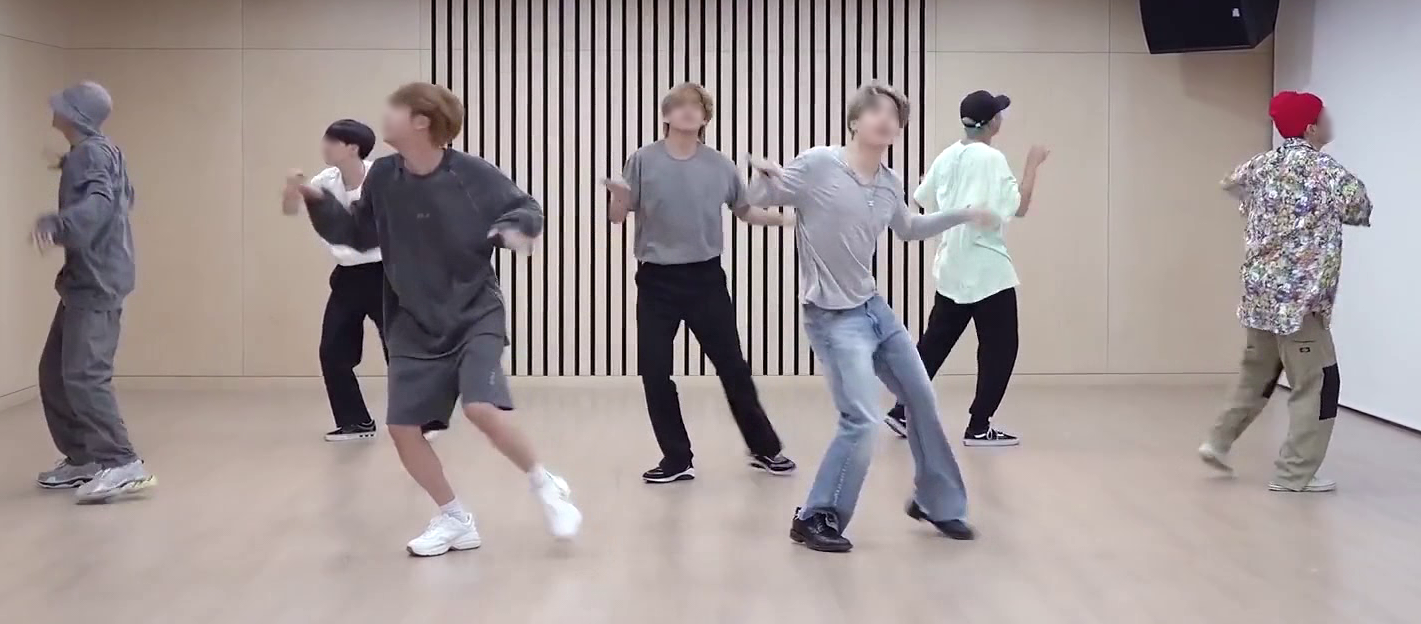}}\\[3pt]
\shortstack{\includegraphics[width=0.33\linewidth]{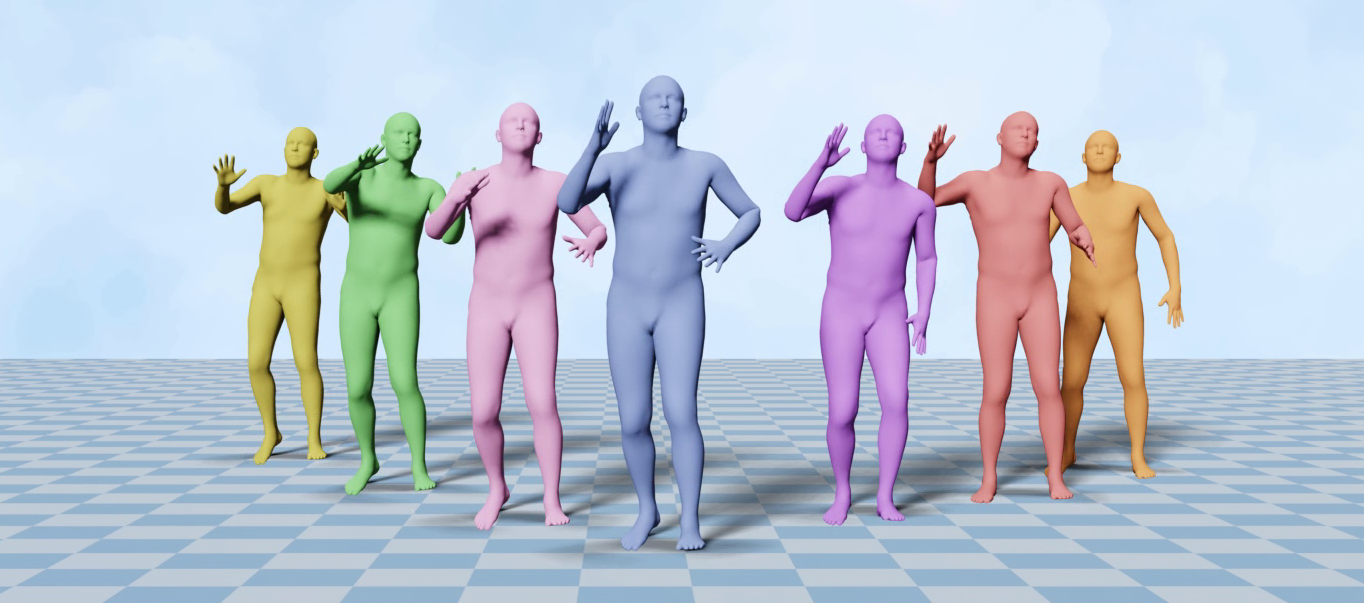}}&
\shortstack{\includegraphics[width=0.33\linewidth]{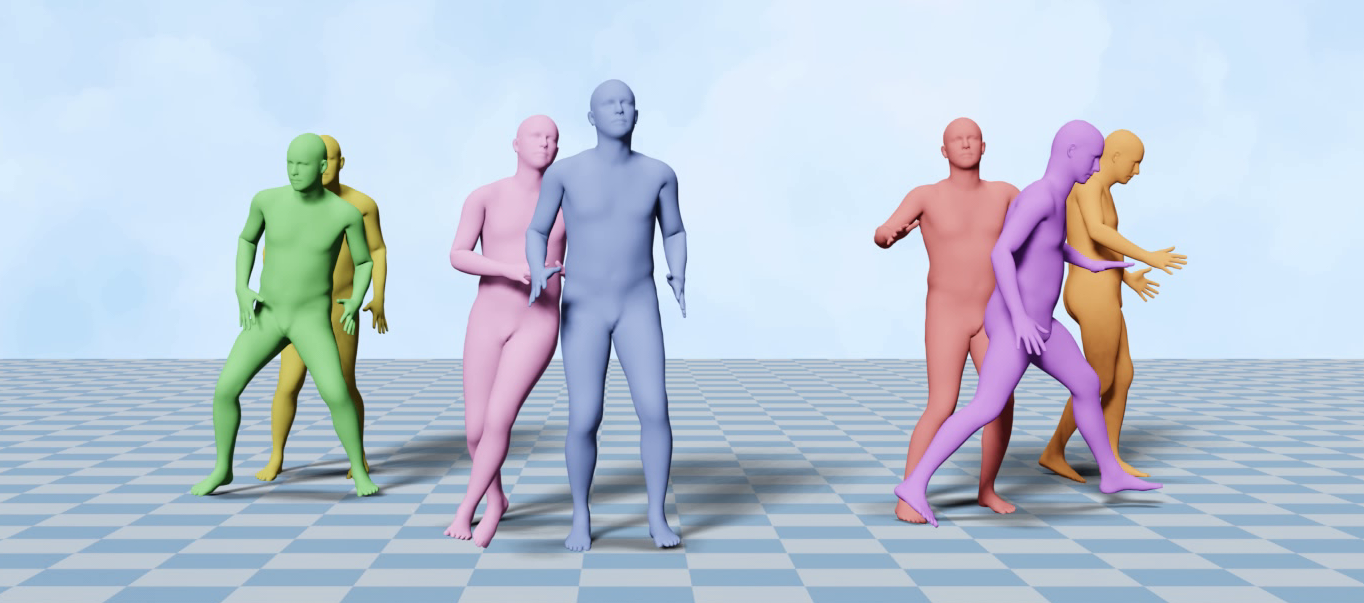}}&
\shortstack{\includegraphics[width=0.33\linewidth]{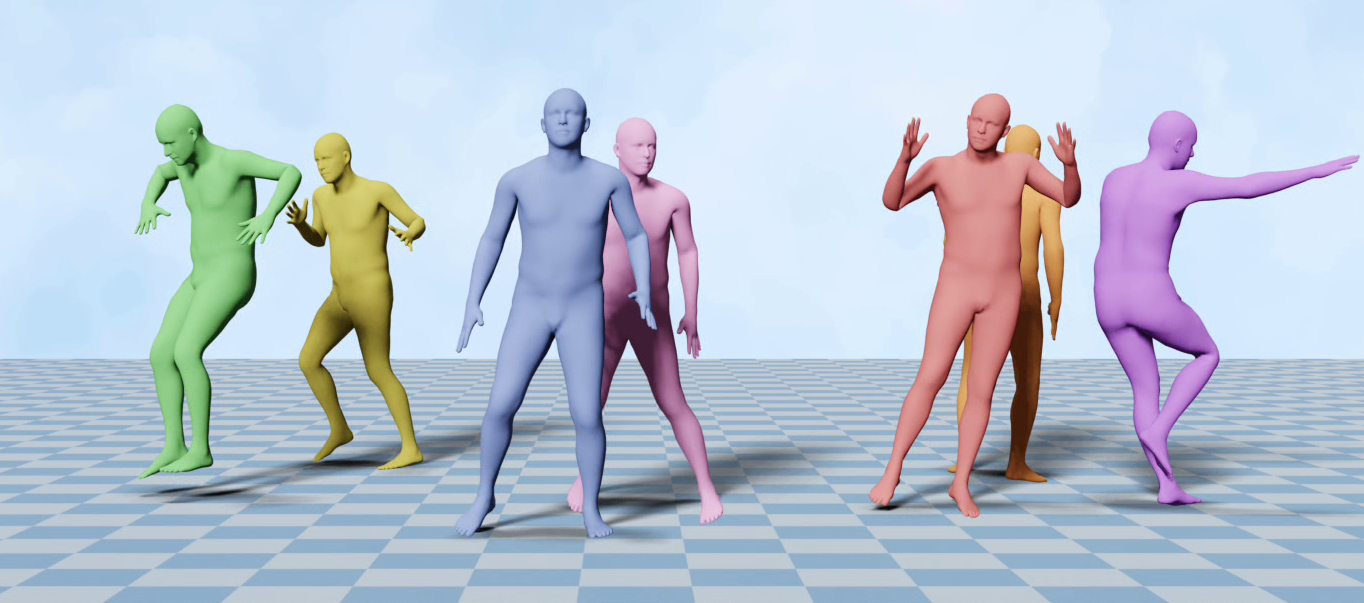}}&
\shortstack{\includegraphics[width=0.33\linewidth]{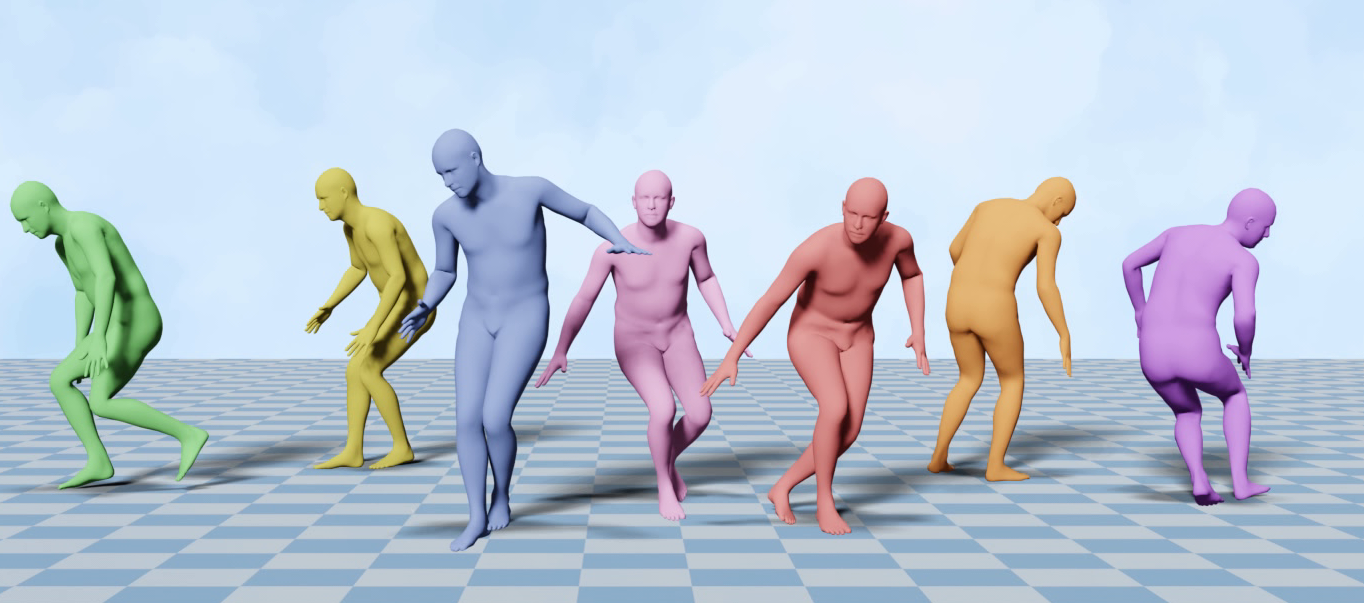}}&
\shortstack{\includegraphics[width=0.33\linewidth]{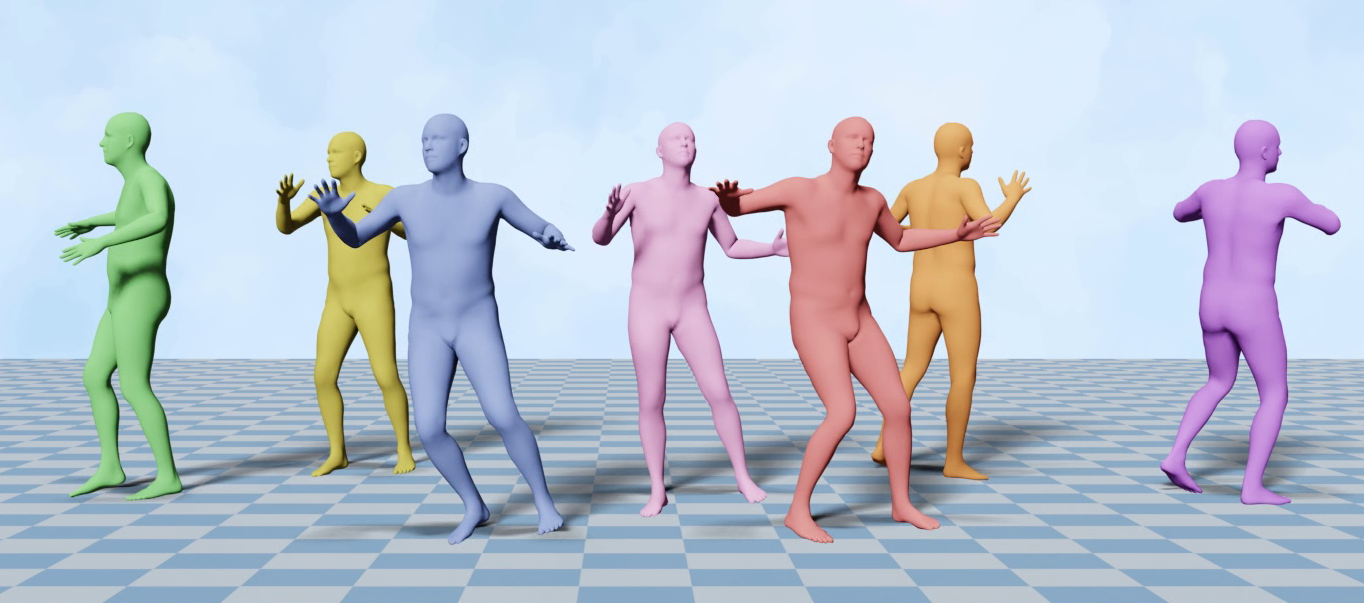}}\\[3pt]
\shortstack{\includegraphics[width=0.33\linewidth]{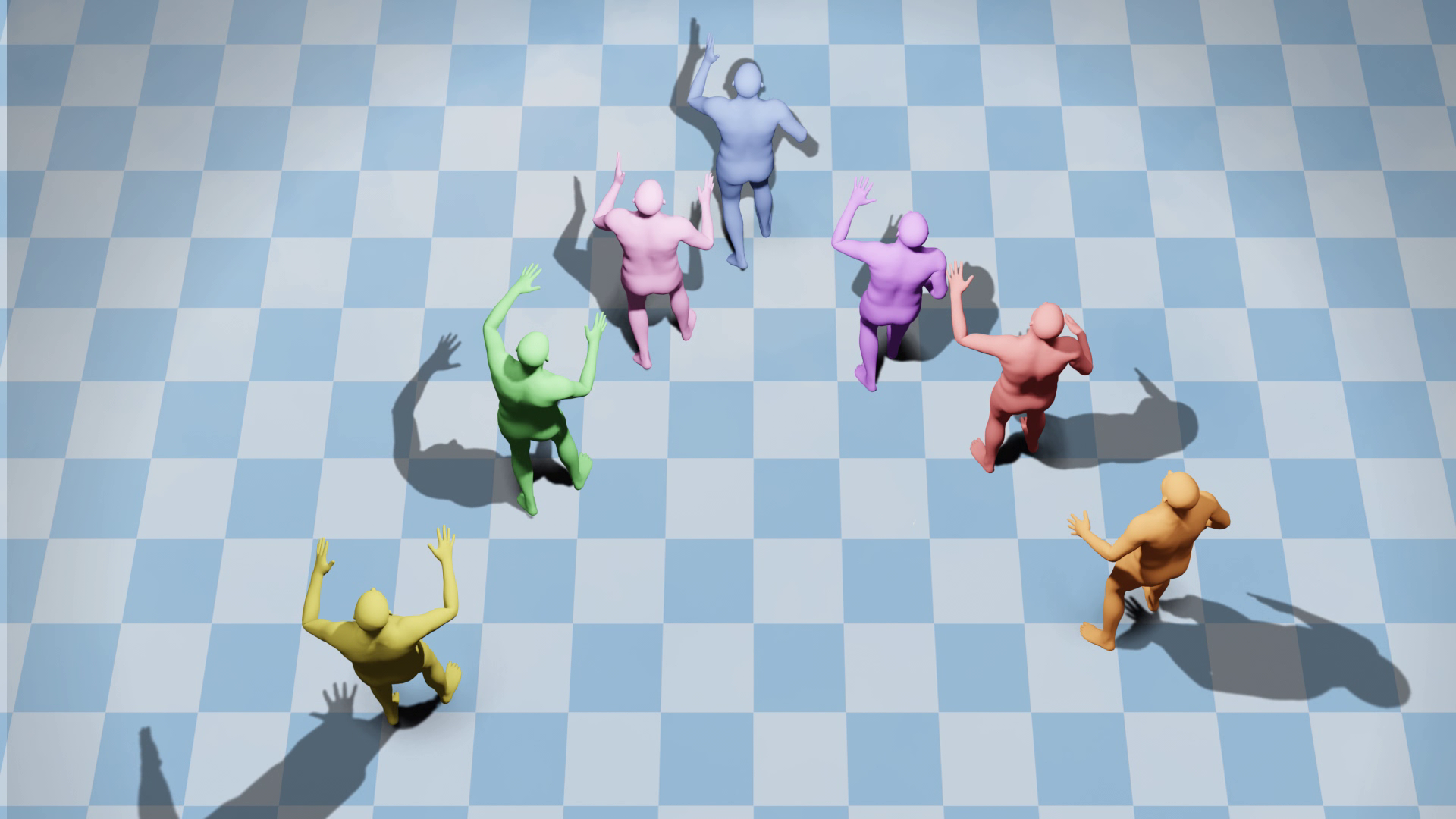}}&
\shortstack{\includegraphics[width=0.33\linewidth]{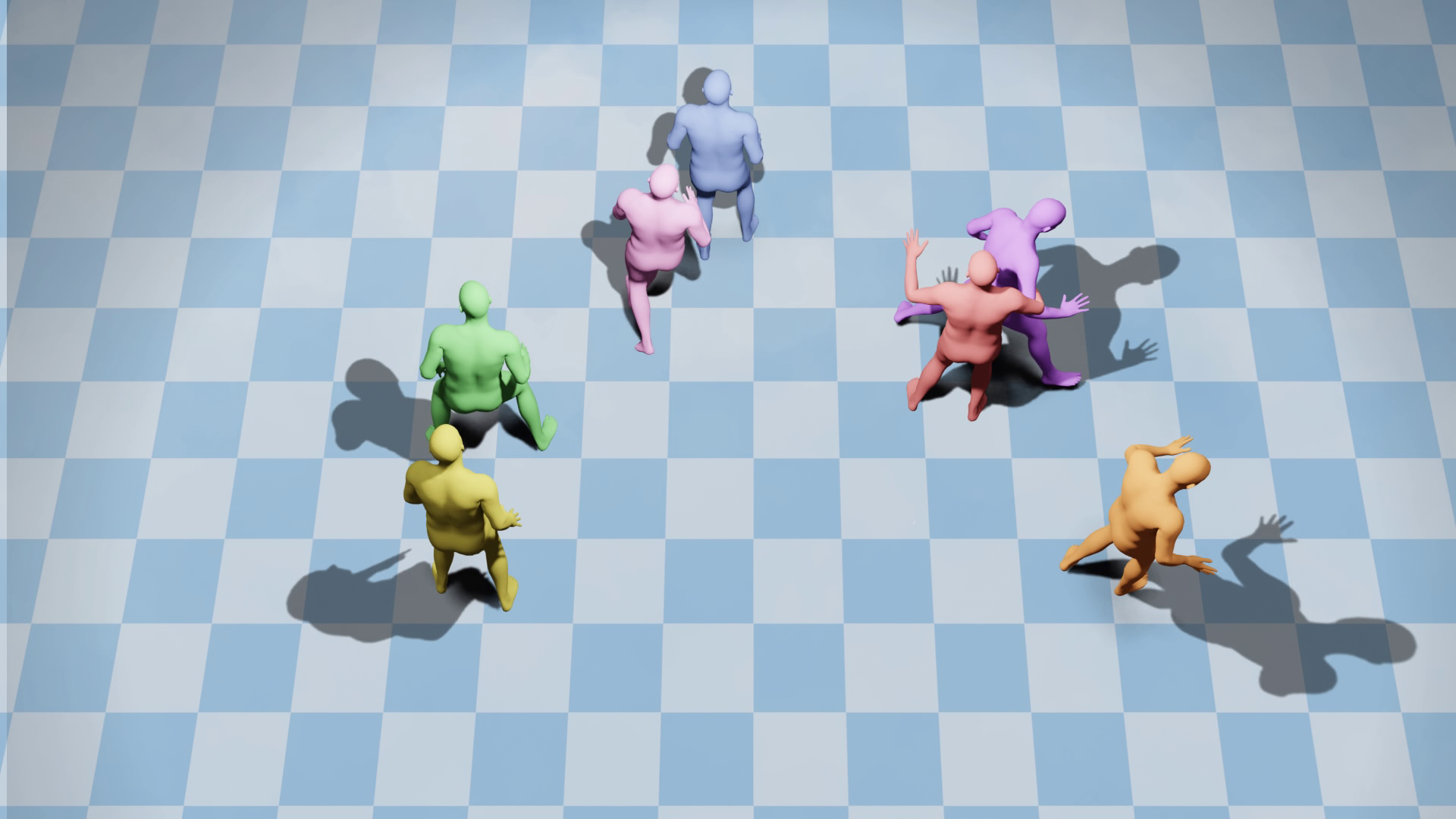}}&
\shortstack{\includegraphics[width=0.33\linewidth]{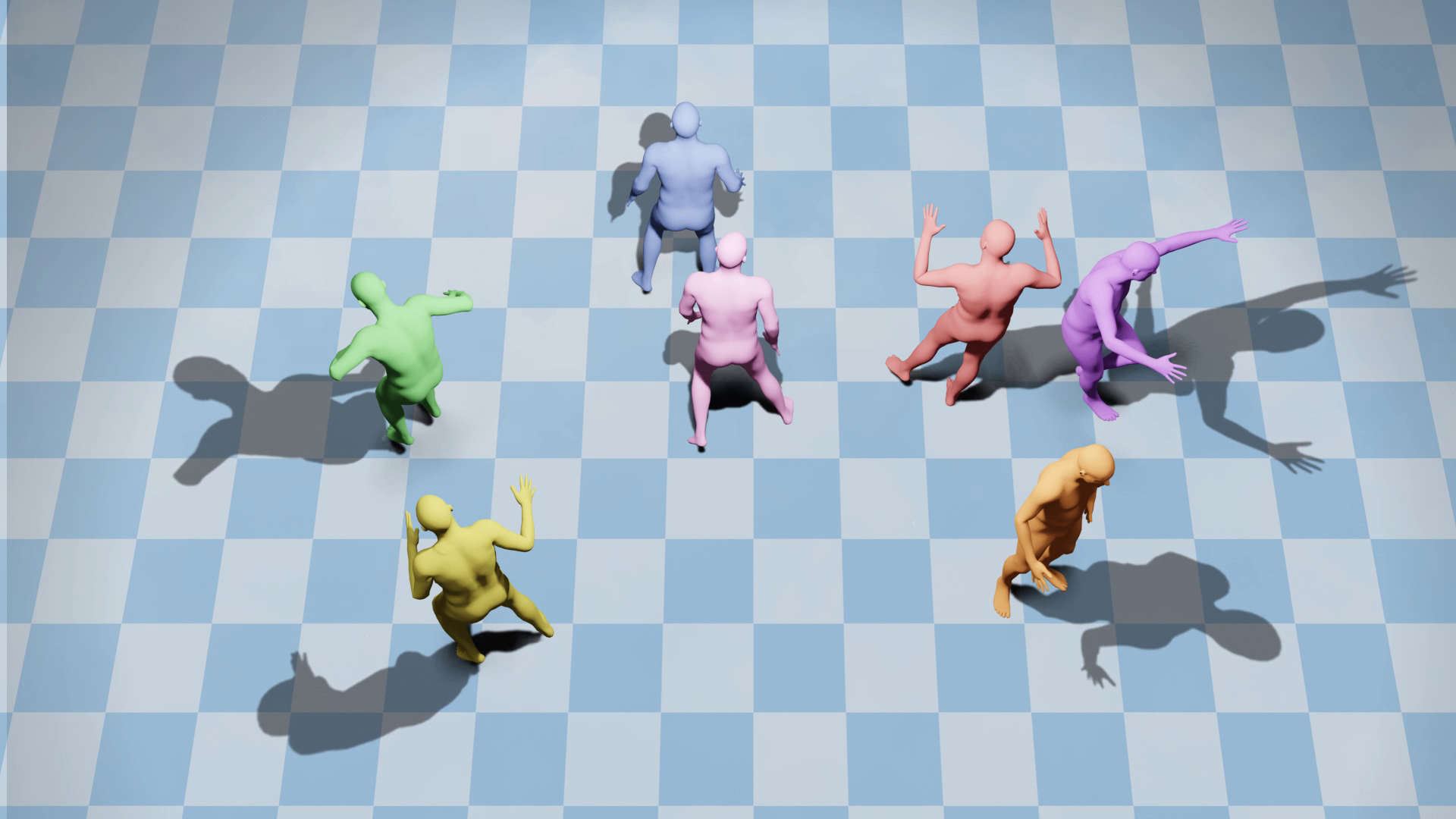}}&
\shortstack{\includegraphics[width=0.33\linewidth]{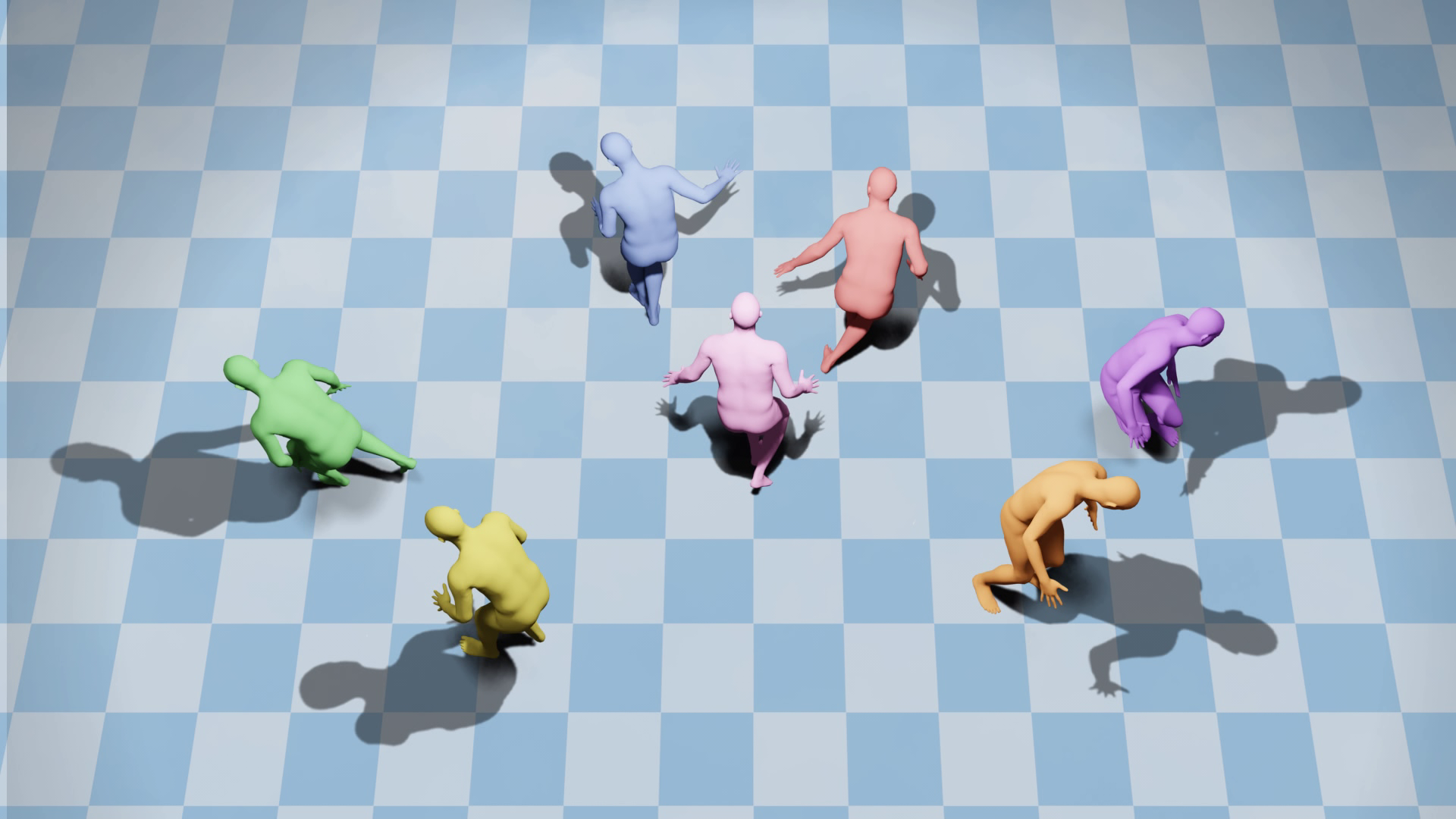}}&
\shortstack{\includegraphics[width=0.33\linewidth]{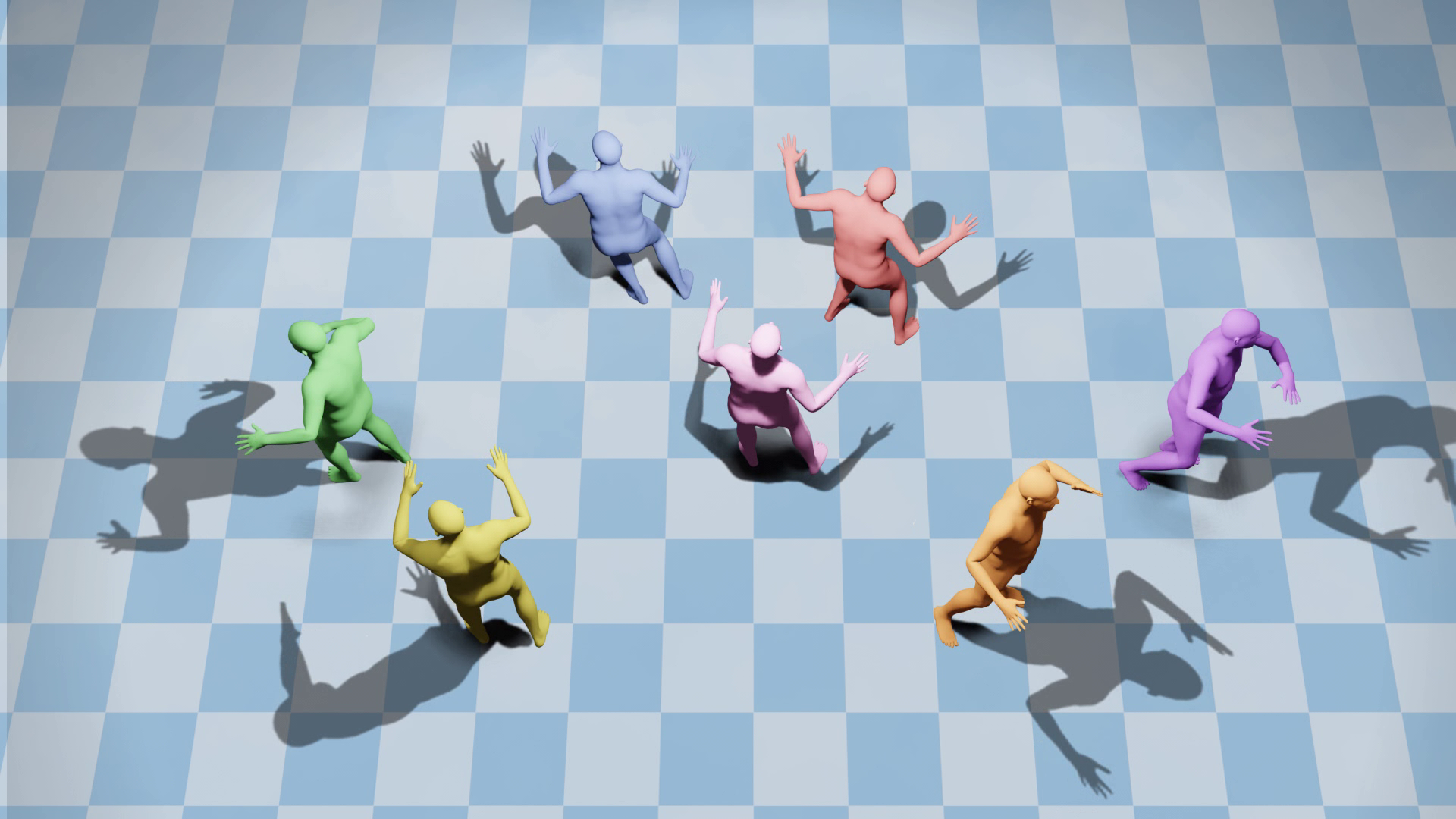}}
\end{tabular}
}
    \caption{Example motion sequence from our dataset from front-view and top-view.}
    \label{fig:Dataset_Sequence}
\end{figure*}

\textbf{Dataset Analysis.} In Figure~\ref{fig:vis_Full_Balanced_Stat}, we show the distribution of music genres and dance styles in our dataset. As illustrated in Figure~\ref{fig_stat_music_genre}, \texttt{Pop} and \texttt{Electronic} are popular music genres while other music genres nearly share the same distribution. In Figure~\ref{fig_stat_dance_style}, \texttt{Zumba}, \texttt{Aerobic}, and \texttt{Commercial} are dominant dance styles. 

Figure~\ref{fig_corr_dancegroup_number} shows the number of dancers in each dance styles. Naturally, we see that \texttt{Zumba}, \texttt{Aerobic}, and \texttt{Commercial} have more dancers. Figure~\ref{fig_corr_dancemusic} shows the correlation between music genres and dance styles. In Figure~\ref{fig:Dataset_Sequence}, we show an example sequence of a dancing motion from our \textsf{AIOZ-GDANCE} dataset.
We recommend the readers to check our Supplementary Material and Demonstration Video for more detailed analysis and illustration.

\subsection{Group Dance Generation Result}

\subsubsection{Implementation Details}

The MLP in Equation~\ref{eq:initial_pose_generator} and~\ref{eq:decoder} has the same architecture with three hidden layers of $512$ neurons each. We apply layer normalization~\cite{ba2016layernorm} and ReLU non-linearity at each hidden layer. The Transformer Music Encoder has 2 transformer layers with 8 attention heads. Both the hidden audio and hidden motion have dimension $d_a = d_h = 1024$. We stack $L=3$ identical Group Encoder layers in the Group Motion Generator to enhance the learning capacity of the model. For the Cross-entity Attention, we also employ multi-head attention strategy with 8 heads and the dimension of query, key, and value for each head is set to $d_k=d_v = 64$. 
During training, we randomly sample the dance motion with the sequence length $T=240$ frames and train the model using $L2$ loss as in \cite{li2021AIST++}. 
We also use scheduled sampling~\cite{bengio2015scheduled} to improve the model robustness and enable long-term generation. The whole model is trained end-to-end using Adam optimizer~\cite{kingma2014adam} with batch size of 16 and learning rate of $10^{-4}$.  At test time, the group dance motions are generated in an auto-regressive manner based on the given inputs.

\subsubsection{Evaluation Protocol}
\label{sec:metric}

We use the following metrics to evaluate the quality of single dancing motion:~ Frechet Inception Distance (FID)~\cite{heusel2017ganfid,li2021AIST++}, Motion-Music Consistency (MMC)~\cite{li2021AIST++}, Generation Diversity (GenDiv)~\cite{Dance_Revolution, lee2019_dancing2music,li2021AIST++}. To evaluate the group dancing quality, we propose three new metrics: Group Motion Realism (GMR), Group Motion Correlation (GMC), and Trajectory Intersection Frequency (TIF). Please see our Supplementary Material for more discussion.

\subsubsection{Experimental Results}
\label{sec:quantitative}
We compare our method with FACT~\cite{li2021AIST++}. FACT is designed for single dance generation, thus giving our method an advantage. However, it is still the closest competing method as we propose a new group dance dataset that is not available for benchmarking before. We also analyse our method with and without using the Cross-entity Attention. We train all methods with mini-batch containing all dancers within the group instead of sampling each dancer independently as in FACT's original implementation.

\textbf{ Cross-entity Attention Analysis.} Table~\ref{tab:baseline} shows the method comparison between the baseline FACT~\cite{li2021AIST++} and our proposed GDanceR with and without Cross-entity Attention. The results show that GDanceR, especially with the Cross-entity Attention, out-performs the baseline by a large margin in all metrics. In Figure~\ref{fig:AblCompare}, we also visualize the example outputs of FACT and GDanceR. It is clear that FACT does not handle well the intersection problem. This is understandable as FACT is not designed for group dance generation, while our method with the Cross-entity Attention can deal with this problem better.

\begin{table}[!t]
\centering
\resizebox{\linewidth}{!}{
\setlength{\tabcolsep}{0.2 em} 
{\renewcommand{\arraystretch}{1.2}
\begin{tabular}{lc|ccc|ccc}
\hline
\multicolumn{2}{l|}{\multirow{2}{*}{{Method}}} & \multicolumn{3}{c|}{{Single-dance Metric}} & \multicolumn{3}{c}{{Group-dance Metric}}  \\ \cline{3-8} 
\multicolumn{2}{l|}{} & \multicolumn{1}{c|}{FID$\downarrow$} & \multicolumn{1}{c|}{MMC$\uparrow$} & GenDiv$\uparrow$ & \multicolumn{1}{c|}{GMR$\downarrow$} & \multicolumn{1}{c|}{GMC$\uparrow$} & TIF$\downarrow$ \\ \hline
\multicolumn{2}{l|}{FACT~\cite{li2021AIST++}} & \multicolumn{1}{c|}{56.20} & \multicolumn{1}{c|}{0.222} & 8.64 & \multicolumn{1}{c|}{101.52} & \multicolumn{1}{c|}{62.68} & 0.321 \\ \hline
\multicolumn{1}{l|}{\multirow{2}{*}{\textbf{\begin{tabular}[c]{@{}c@{}}GDanceR \\ (ours)\end{tabular}}}} & \textit{w/o} CA & \multicolumn{1}{c|}{63.83} & \multicolumn{1}{c|}{0.218} & 8.99 & \multicolumn{1}{c|}{109.80} & \multicolumn{1}{c|}{68.47} & 0.379\\ 
\cline{2-8} 
\multicolumn{1}{l|}{} & \textbf{\textit{w} CA} & \multicolumn{1}{c|}{\textbf{43.90}} & \multicolumn{1}{c|}{\textbf{0.250}} & \textbf{9.23} & \multicolumn{1}{c|}{\textbf{51.27}} & \multicolumn{1}{c|}{\textbf{79.01}} & \textbf{0.217}\\ \hline
\end{tabular}
}}
\caption{The generation results on our dataset. \textit{w/o} CA denotes without using Cross-entity Attention.}
    \label{tab:baseline}
\end{table}

\begin{figure}[ht] 
  \centering
  \footnotesize
\resizebox{\linewidth}{!}{
\setlength{\tabcolsep}{2pt}
\begin{tabular}{ccc}
\shortstack{\includegraphics[width=0.33\linewidth]{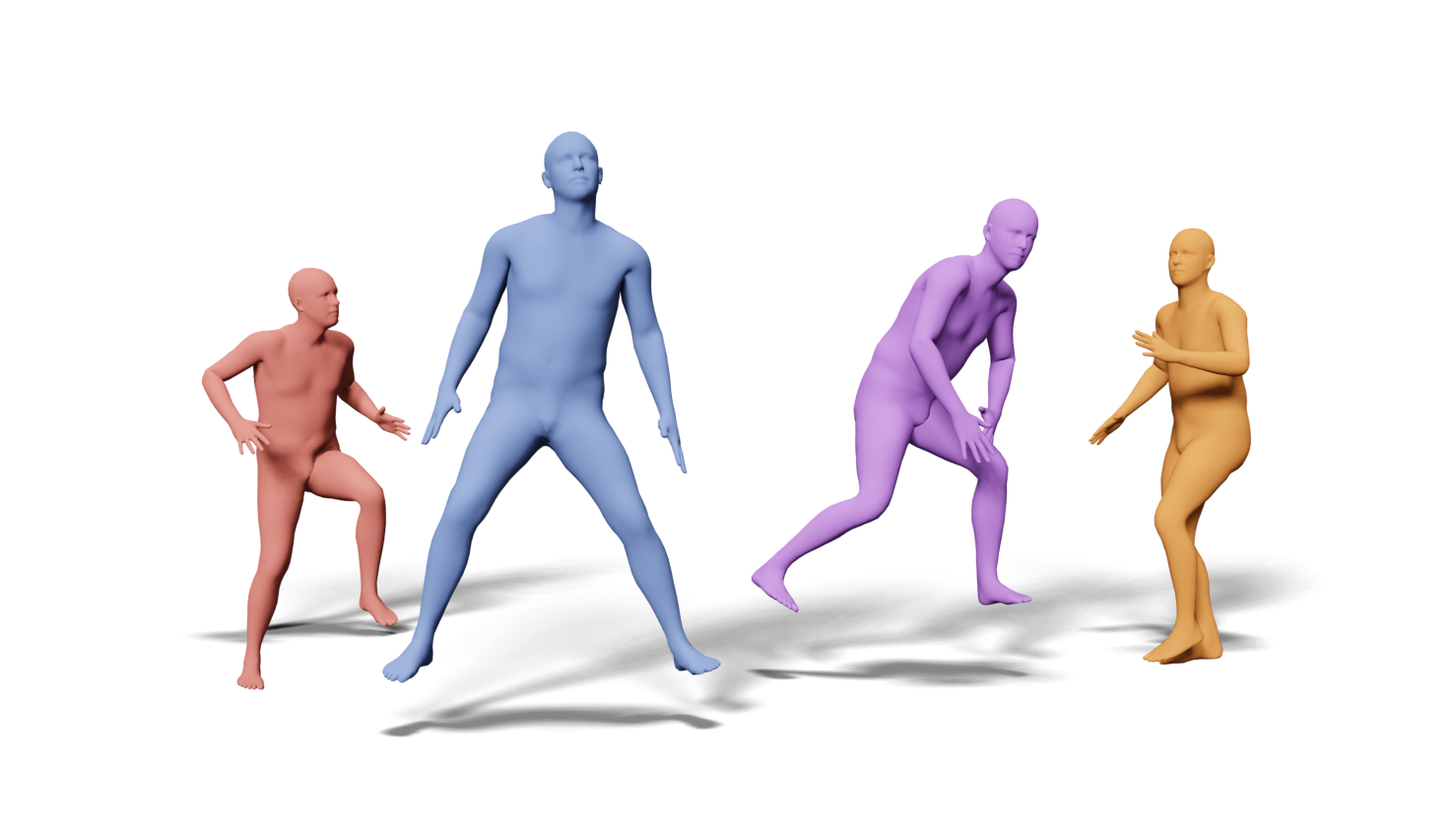}}&
\shortstack{\includegraphics[width=0.33\linewidth]{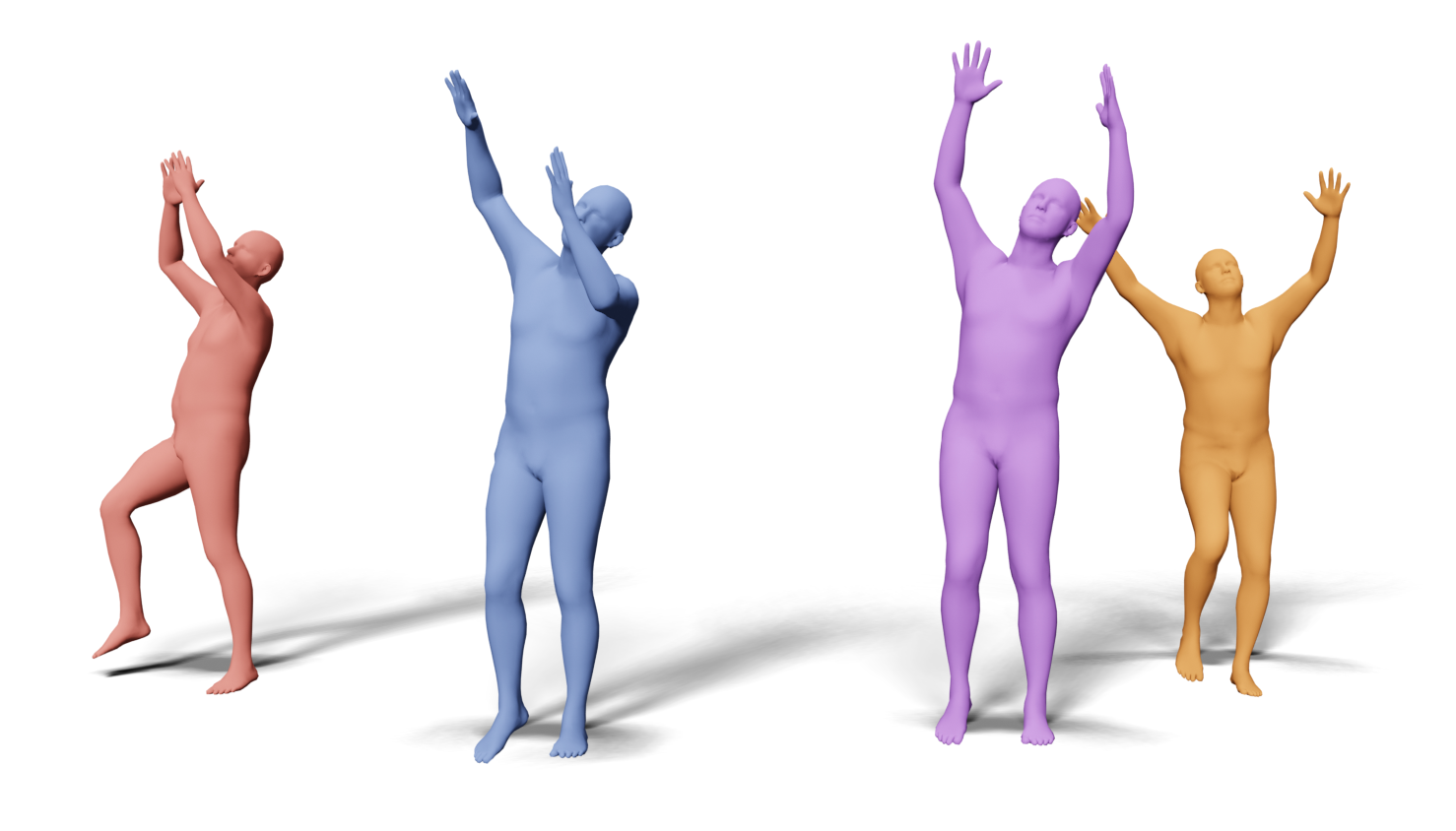}}\\[3pt]
\shortstack{\includegraphics[width=0.33\linewidth]{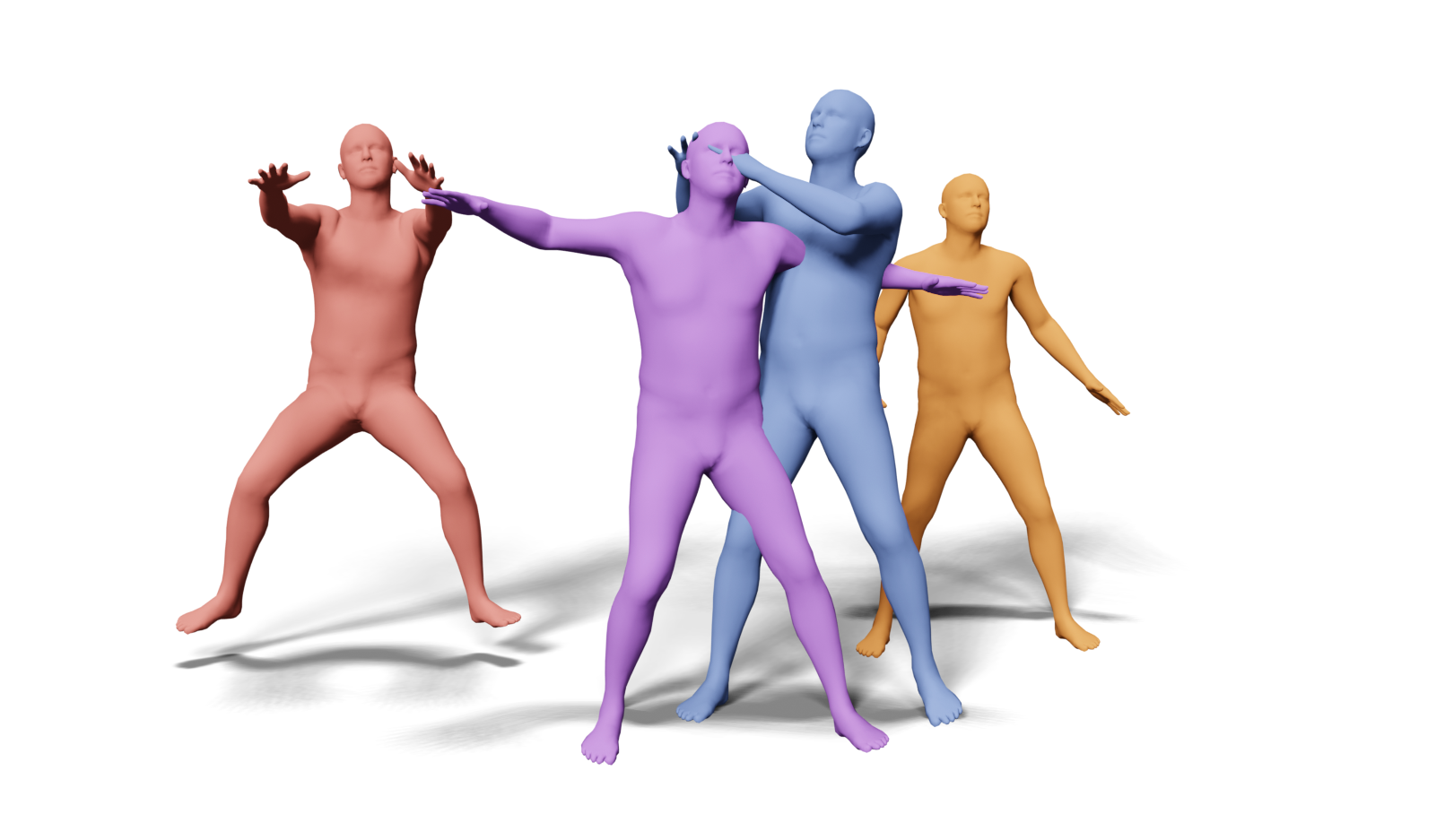}}&
\shortstack{\includegraphics[width=0.33\linewidth]{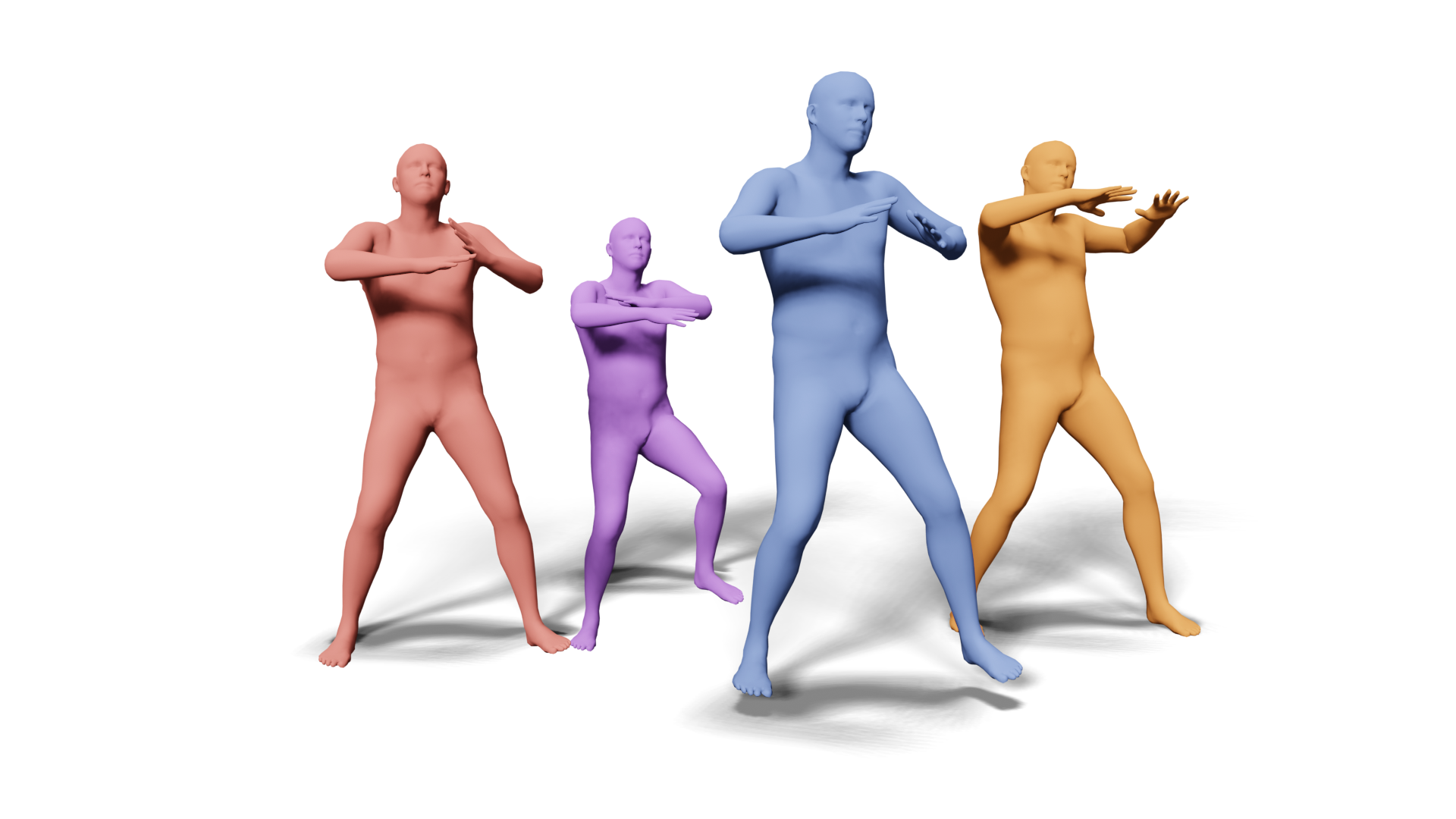}}\\[3pt]
\shortstack{\scriptsize (a) FACT~\cite{li2021AIST++}}&
\shortstack{\scriptsize (b) GDanceR}
\end{tabular}
}
\vspace{-2ex}
    \caption{Comparison between FACT~\cite{li2021AIST++} and our GDanceR. Our method handles better the consistency and cross-body intersection problem between dancers.}
    \label{fig:AblCompare}
\end{figure}

\textbf{Number of Dancers Analysis.}
Table~\ref{tab:n_dancers} demonstrates the generation results of our method when we want to generate different numbers of dancers. In general, the FID, GMR, and GMC metric do not show much correlation with the numbers of generated dancers since the results are varied. On the other hand, MMC shows its stability among all setups ($\sim 0.248$), which indicates that our network is robust in generating motion from given music regardless of the changing of initial positions. The generation diversity (GenDiv) decreases while the intersection frequency (TIF) increases when more dancers are generated. These results show that dealing with the collision during the group generation process is worth further investigation. 

\begin{table}[!t]
\centering
\resizebox{\linewidth}{!}{
\setlength{\tabcolsep}{0.3 em} 
{\renewcommand{\arraystretch}{1.2}
\begin{tabular}{c|ccc|ccc}
\hline
\multirow{2}{*}{{\begin{tabular}[c]{@{}c@{}}$N$ Generated\\ Dancers\end{tabular}}} & \multicolumn{3}{c|}{{Single-dance Metric}} & \multicolumn{3}{c}{{Group-dance Metric}} \\ \cline{2-7} 
 & \multicolumn{1}{c|}{FID$\downarrow$} & \multicolumn{1}{c|}{MMC$\uparrow$} & GenDiv$\uparrow$ & \multicolumn{1}{c|}{GMR$\downarrow$} & \multicolumn{1}{c|}{GMC$\uparrow$} & TIF$\downarrow$ \\ \hline
\textit{2} & \multicolumn{1}{c|}{48.82} & \multicolumn{1}{c|}{0.248} & 9.66 & \multicolumn{1}{c|}{53.83} & \multicolumn{1}{c|}{75.44} & 0.086 \\ 
\textit{3} & \multicolumn{1}{c|}{44.47} & \multicolumn{1}{c|}{0.245} & 9.46 & \multicolumn{1}{c|}{52.85} & \multicolumn{1}{c|}{74.07} & 0.104 \\ 
\textit{4} & \multicolumn{1}{c|}{47.32} & \multicolumn{1}{c|}{0.248} & 9.24 & \multicolumn{1}{c|}{58.79} & \multicolumn{1}{c|}{77.71} & 0.162 \\ 
\textit{5} & \multicolumn{1}{c|}{44.19} & \multicolumn{1}{c|}{0.249} & 9.38 & \multicolumn{1}{c|}{55.05} & \multicolumn{1}{c|}{78.72} & 0.218 \\ 
\textit{6} & \multicolumn{1}{c|}{50.95} & \multicolumn{1}{c|}{0.250} & 9.25 & \multicolumn{1}{c|}{59.05} & \multicolumn{1}{c|}{75.24} & 0.319 \\ 
\textit{7} & \multicolumn{1}{c|}{48.86} & \multicolumn{1}{c|}{0.250} & 9.19 & \multicolumn{1}{c|}{56.23} & \multicolumn{1}{c|}{76.01} & 0.367 \\ \hline
\end{tabular}
}}
\caption{Performance of our proposed method when we increase the number of generated dancers.}
    \label{tab:n_dancers}
\vspace{-1ex}
\end{table}

\begin{figure}[ht] 

\setlength{\tabcolsep}{2 pt}
\begin{tabular}{ccccccc}
\shortstack{\includegraphics[width=0.32\linewidth]{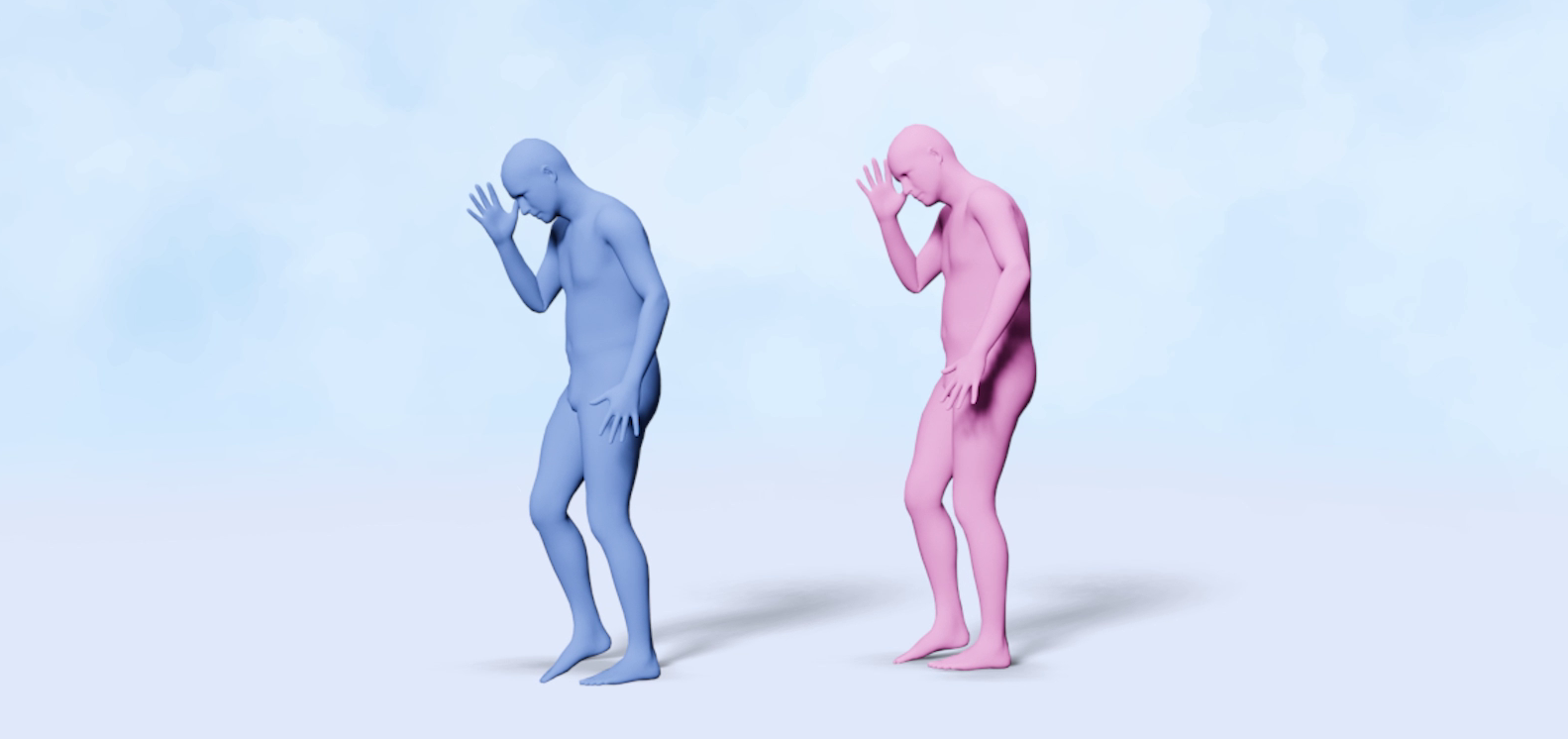}}&
\shortstack{\includegraphics[width=0.32\linewidth]{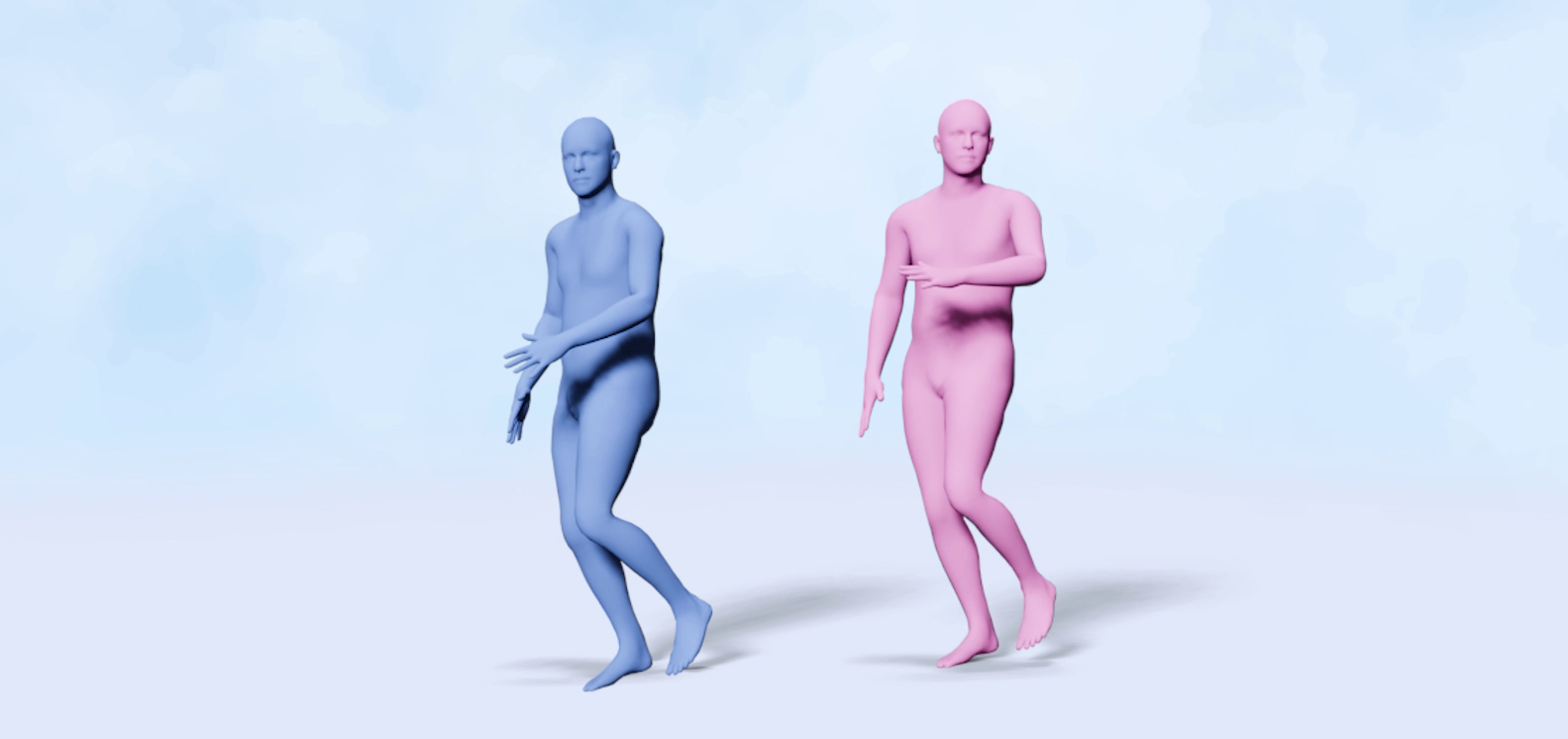}}&
\shortstack{\includegraphics[width=0.32\linewidth]{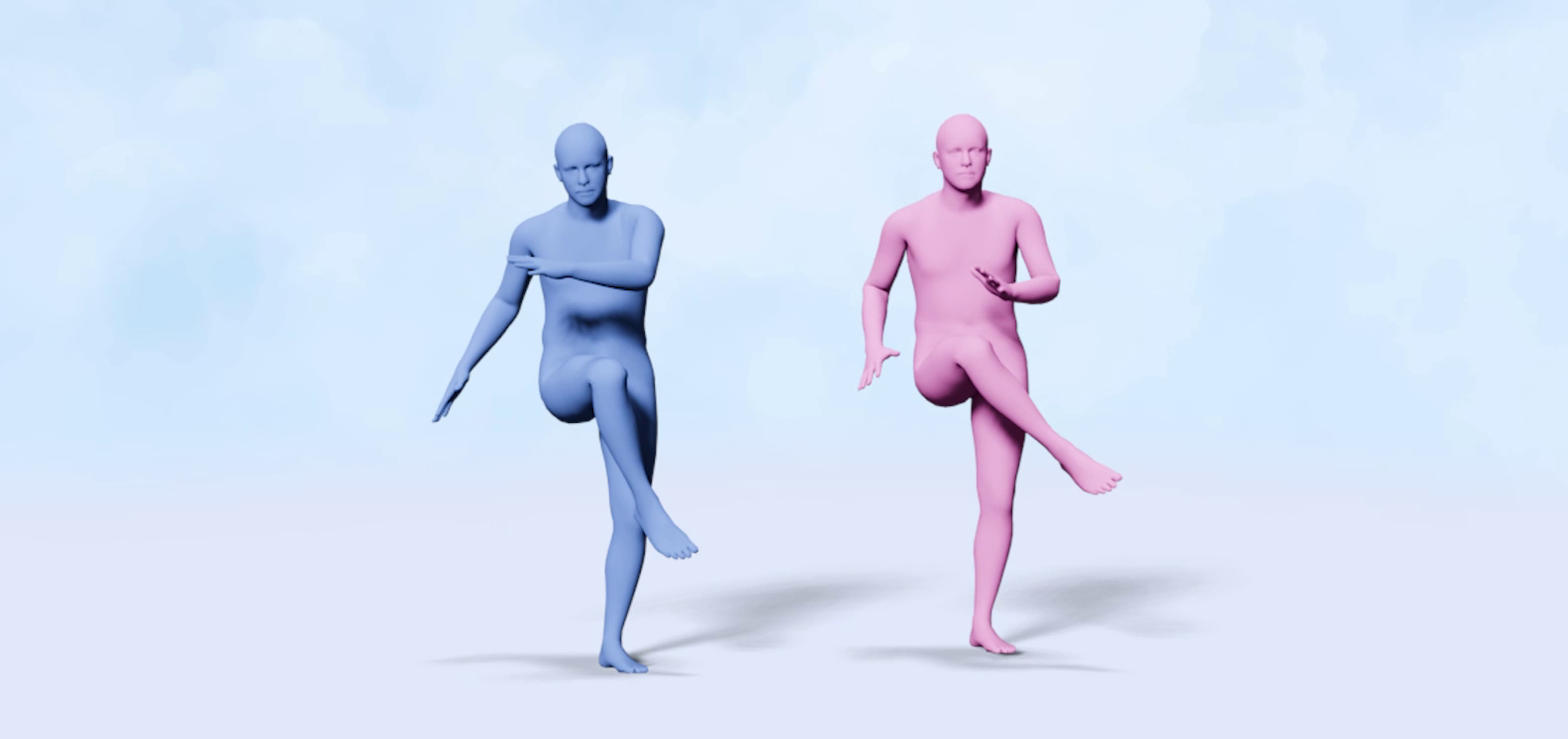}}\\[3pt]
\shortstack{\includegraphics[width=0.32\linewidth]{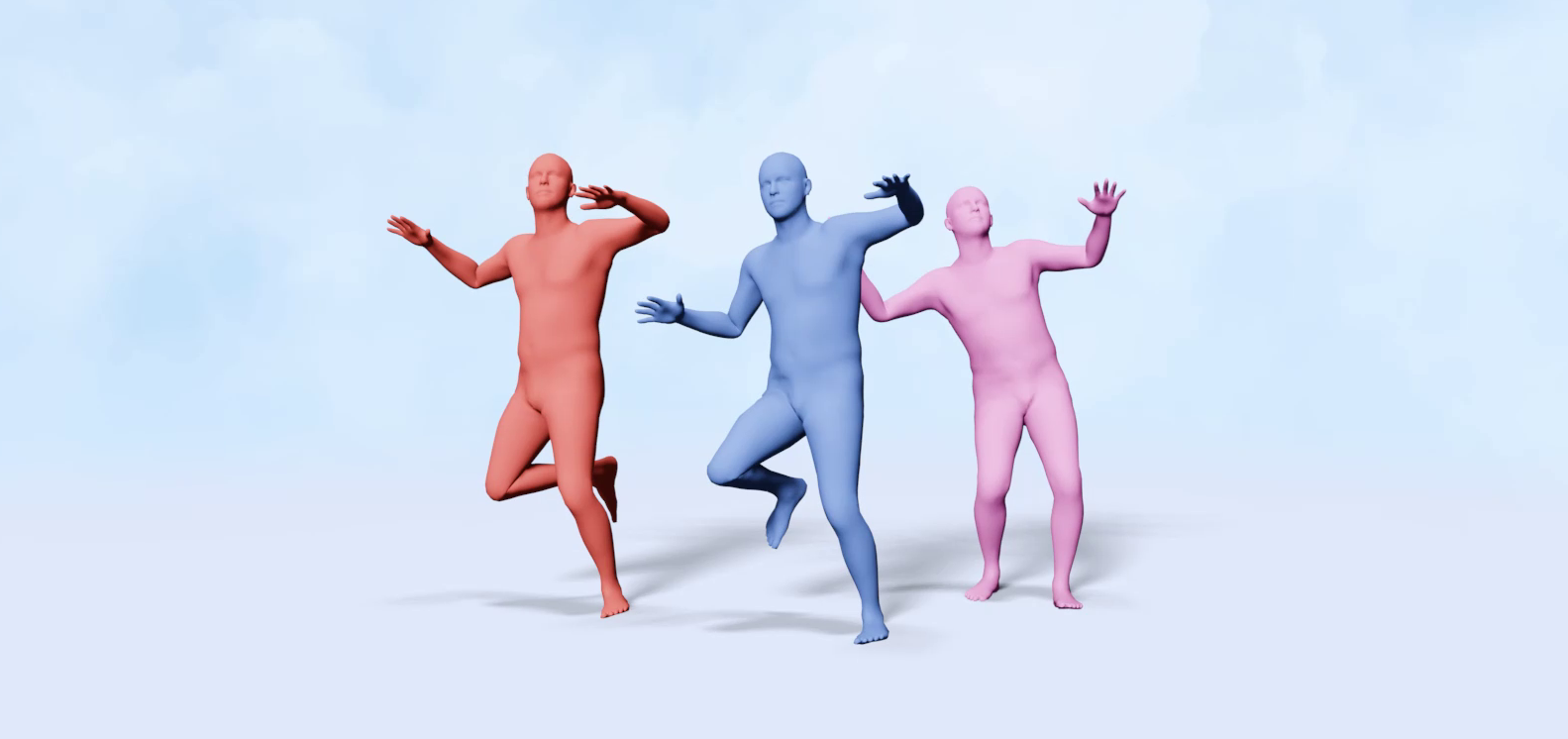}}&
\shortstack{\includegraphics[width=0.32\linewidth]{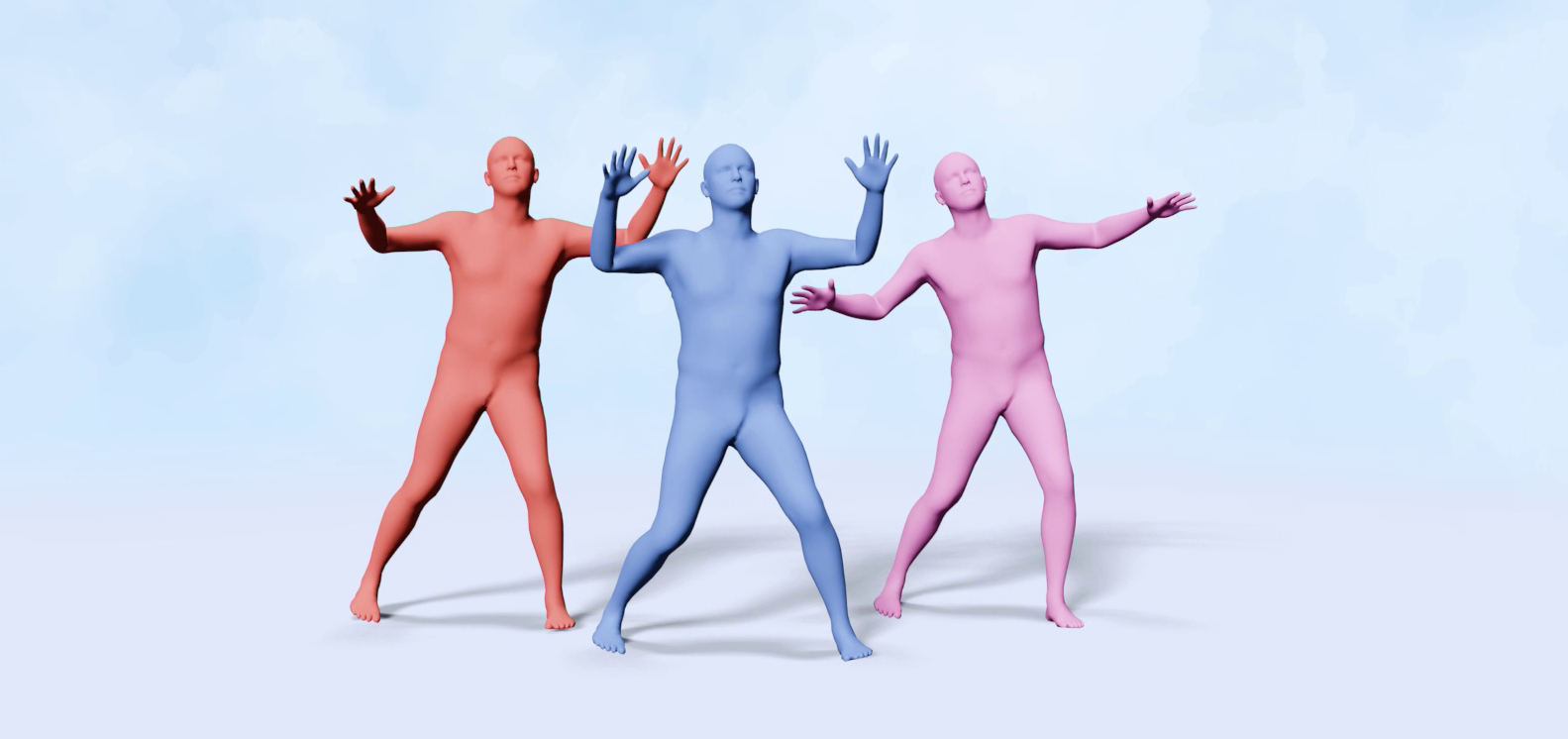}}&
\shortstack{\includegraphics[width=0.32\linewidth]{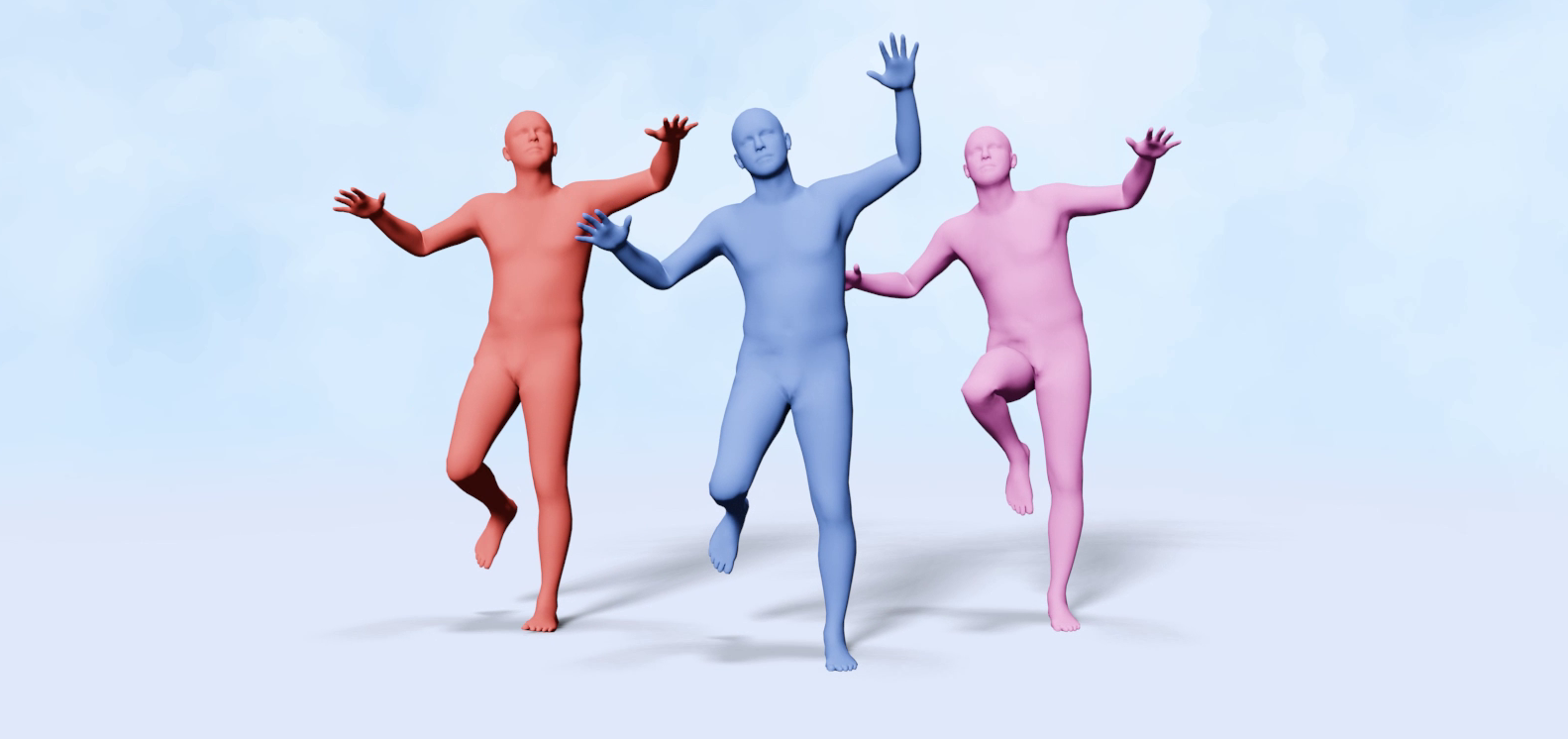}}\\[3pt]
\shortstack{\includegraphics[width=0.32\linewidth]{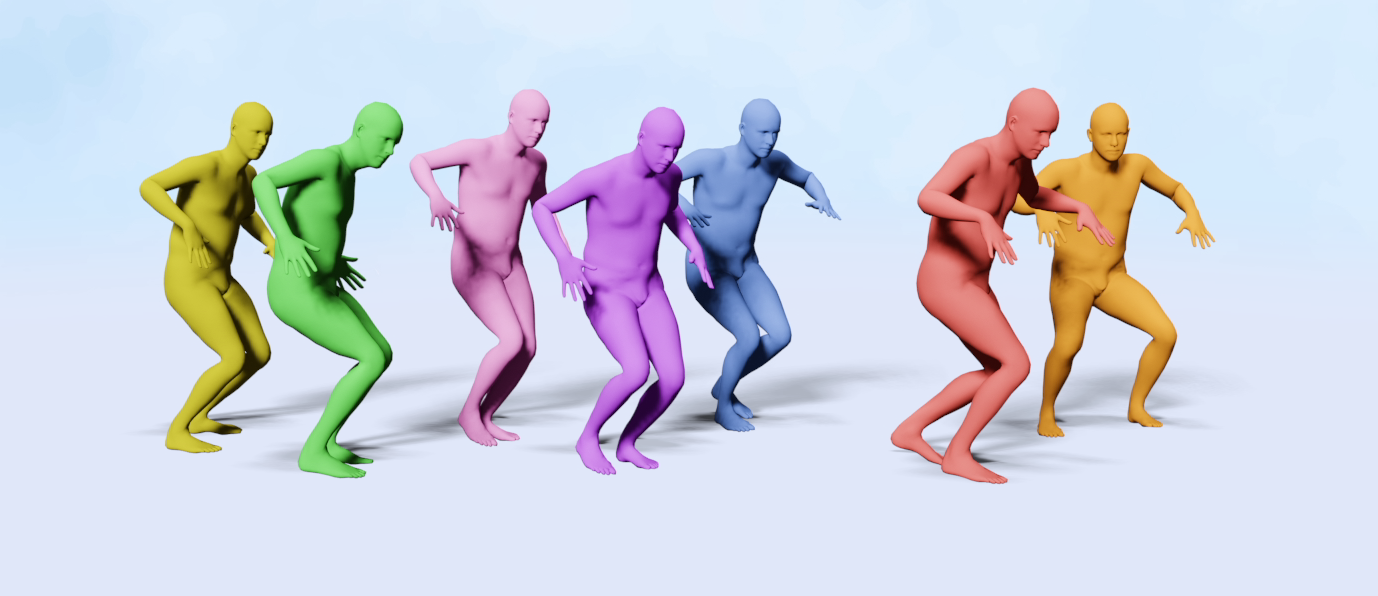}}&
\shortstack{\includegraphics[width=0.32\linewidth]{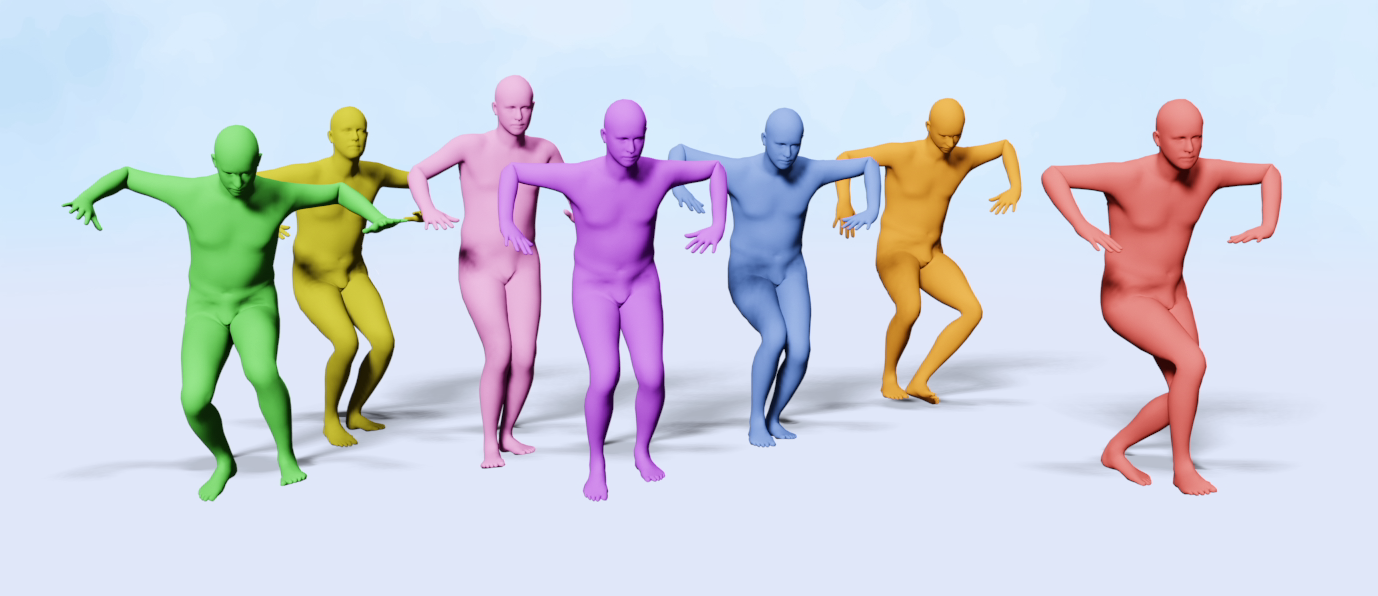}}&
\shortstack{\includegraphics[width=0.32\linewidth]{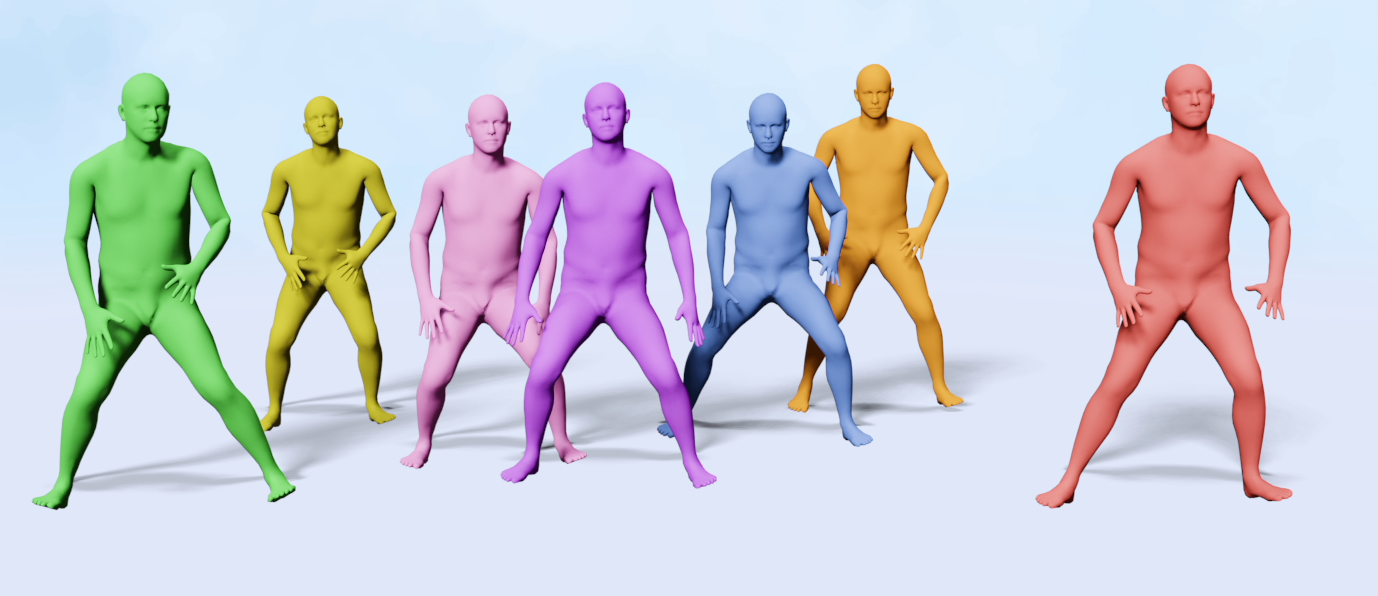}
}
\end{tabular}
\vspace{-1ex}
    \caption{Examples of generated group motions from our method.} 
    \label{fig:result_Vis}
\end{figure}

\textbf{Dance Style Analysis.} Different dance styles exhibit different challenges in group dance generation. 
As shown in Table~\ref{tab:danceStyle_baseline}, \texttt{Aerobic} and \texttt{Zumba} are quite similar for generating choreography as they usually focus on workout and sporty movements.
Besides, while \texttt{Commercial} and \texttt{Irish} are easier for the model to generate, \texttt{Bollywood} and \texttt{Samba} contain highly skilled movements that are challenging to capture and represent accurately. In Figure~\ref{fig:result_Vis}, we show the generated results of GDanceR with different dance styles. Our Supplementary Material and Demonstration Video also provide more examples.

\begin{table}[!t]
\centering
\resizebox{\linewidth}{!}{
\setlength{\tabcolsep}{0.3 em}
{\renewcommand{\arraystretch}{1.2}
\begin{tabular}{c|ccc|ccc}
\hline
\multirow{2}{*}{\begin{tabular}[c]{@{}c@{}}Dance Styles\end{tabular}} & \multicolumn{3}{c|}{{Single-dance Metric}} & \multicolumn{3}{c}{{Group-dance Metric}} \\ \cline{2-7} 
  & \multicolumn{1}{c|}{FID$\downarrow$} & \multicolumn{1}{c|}{MMC$\uparrow$} & GenDiv$\uparrow$ & \multicolumn{1}{c|}{GMR$\downarrow$} & \multicolumn{1}{c|}{GMC$\uparrow$} & TIF$\downarrow$ \\ \hline
\texttt{Zumba} & \multicolumn{1}{c|}{45.86} & \multicolumn{1}{c|}{0.268} & 9.77 & \multicolumn{1}{c|}{50.97} & \multicolumn{1}{c|}{72.70} & 0.133 \\ 
\texttt{Aerobic} & \multicolumn{1}{c|}{38.68} & \multicolumn{1}{c|}{0.252} & 6.57 & \multicolumn{1}{c|}{63.62} & \multicolumn{1}{c|}{75.12} & 0.249 \\ 
\texttt{Commercial} & \multicolumn{1}{c|}{46.22} & \multicolumn{1}{c|}{0.232} & 8.58 & \multicolumn{1}{c|}{51.18} & \multicolumn{1}{c|}{81.02} & 0.056 \\ 
\texttt{Bollywood} & \multicolumn{1}{c|}{81.89} & \multicolumn{1}{c|}{0.211} & 2.14 & \multicolumn{1}{c|}{101.49} & \multicolumn{1}{c|}{74.00} & 0.377\\ 
\texttt{Irish} & \multicolumn{1}{c|}{42.02} & \multicolumn{1}{c|}{0.219} & 8.56 & \multicolumn{1}{c|}{42.73} & \multicolumn{1}{c|}{82.00} & 0.083\\
\texttt{Rumba} & \multicolumn{1}{c|}{69.62} & \multicolumn{1}{c|}{0.273} & 3.91 & \multicolumn{1}{c|}{68.00} & \multicolumn{1}{c|}{71.85} & 0.228\\ 
\texttt{Samba} & \multicolumn{1}{c|}{71.00} & \multicolumn{1}{c|}{0.228} & 7.77 & \multicolumn{1}{c|}{98.83} & \multicolumn{1}{c|}{67.76} & 0.441\\ \hline
\end{tabular}
}}
\caption{The results of different dance styles. Note that these results are obtained by training the model on each dance style.}
    \label{tab:danceStyle_baseline}
\end{table}

\begin{table}
\centering

\resizebox{\linewidth}{!}{
\setlength{\tabcolsep}{0.3 em} 
{\renewcommand{\arraystretch}{1.2}
\begin{tabular}{c|c|c|c|c|c|c} 
\hline
\multirow{2}{*}{\begin{tabular}[c]{@{}c@{}}Fusion\\Strategy\end{tabular}} & \multicolumn{3}{c|}{{Single-dance Metric}} & \multicolumn{3}{c}{{Group-dance Metric}} \\ 
\cline{2-7}
 & FID$\downarrow$ & MMC$\uparrow$  & GenDiv$\uparrow$ & GMR$\downarrow$ & GMC$\uparrow$ & TIF$\downarrow$ \\ 
\hline
No Fusion & 47.19 & 0.242 & 9.14 & 57.84  & 69.67 & 0.221 \\ 

Concatenate & 52.25 & 0.223 & 9.23 & 54.23 & 72.46 & 0.242 \\ 

\textbf{Add} & \textbf{43.90} & \textbf{0.250} & \textbf{9.23}  & \textbf{51.27} & \textbf{79.01} & \textbf{0.217} \\
\hline
\end{tabular}}}
\caption{Ablation study on different fusion strategies for the motion representation.}
\label{tab:motion_fusion}
\vspace{-2ex}
\end{table}

\textbf{Latent Motion Fusion Analysis.} We investigate different fusion strategies between the local motion $h^i$ and global-aware motion $g^i$ to obtain the final motion representation $z^i$. Specifically, we experiment with three settings: \textit{(i)} No Fusion: the final motion is the global-aware motion obtained from our Cross-entity Attention ($z^i = g^i$); \textit{(ii)} Concatenate: the final motion is the concatenation of the local and global-aware motion ($z^i = [h^i; g^i]$); \textit{(iii)} Add: the final motion is the addition between local and global ($z^i = h^i + g^i$). Table~\ref{tab:motion_fusion} summarizes the results. We find that fusing the motion by adding both the local and global motion features achieves the best results.

\section{Conclusion}
We have introduced \textsf{AIOZ-GDANCE}, the largest dataset for audio-driven group dance generation. Our dataset contains in-the-wild videos and covers different dance styles and music genres. We then propose a strong baseline along with evaluation protocols for group dance generation task. We hope that the release of our dataset will foster more research on audio-driven group choreography.

{\small
\bibliographystyle{ieee_fullname}
\bibliography{egbib}
}

\end{document}